\documentclass[fleqn,numbers,english,sort&compress,3p,times]{elsarticle}
\usepackage[T1]{fontenc}
\usepackage[latin9]{inputenc}
\usepackage{geometry}
\usepackage[colorlinks=true]{hyperref}
\hypersetup{citecolor=blue,linkcolor=blue,urlcolor=blue,breaklinks=true}
\usepackage{float}
\usepackage{amsmath}
\usepackage{amssymb}
\usepackage{cancel}
\usepackage{graphicx}
\usepackage{wasysym}
\journal{Nuclear Physics B}

\makeatletter
\@ifundefined{showcaptionsetup}{}{%
 \PassOptionsToPackage{caption=false}{subfig}}
\usepackage{subfig}
\makeatother

\usepackage{babel}
\begin{document}

\begin{frontmatter}{}

\title{The influence of a conducting surface on the conductivity of graphene}

\author{Danilo C. Pedrelli\corref{cor1}} \ead{danilo.pedrelli@icen.ufpa.br}
\author{Danilo T. Alves} \ead{danilo@ufpa.br}
\author{Van Sérgio Alves} \ead{vansergi@ufpa.br}

\address{Faculdade de Física, Universidade Federal do Pará, 66075-110, Belém, Pará, Brazil}

\cortext[cor1]{Corresponding author}

%%%%%%%%%%%%%%%%%%%%%%%%%%%%%%%%%%%%%%%%%%%%%%%%%%%%%%%%%%%%%
\begin{abstract}
In the present paper, using Pseudo-Quantum Electrodynamics to describe the interaction
between electrons in graphene, we investigate the longitudinal and optical conductivities of a neutral graphene sheet near a 
grounded perfectly conducting surface, with calculations up to 2-loop perturbation order.
We show that the longitudinal conductivity increases as we bring the conducting surface closer to the graphene sheet. 
On the other hand, although the optical conductivity initially increases with the proximity of the plate, it reaches a maximum value, tending, afterwards, to the minimal conductivity in the ideal limit of no separation between graphene and the conducting surface.
We recover the correspondent results in the literature when the distance to the plate tends to infinity.
Our results may be useful as an alternative way to control the longitudinal and optical conductivities of graphene.
\end{abstract}

\end{frontmatter}{}

%%%%%%%%%%%%%%%%%%%%%%%%%%%%%%%%%%%%%%%%%%%%%%%%%%%%%%%%%%%%%%%%%%%%%%%%%%%%%%%%%%%%%%%%%%%%%%
\section{Introduction}
\label{sec-introduction}
Pseudo-Quantum Electrodynamics (PQED) describes electrons moving on a plane in a $(2+1)$ dimensional spacetime.
In PQED, although electrons are confined to a plane, they interact, effectively, as particles in $3+1$ dimensions, which leads to a Coulombian interaction potential between static charges, instead of the logarithmic one predicted by the $2 +1$ dimensional Quantum Electrodynamics (QED) \cite{roberts1994dyson}.
The PQED effective Lagrangian is ($\hbar=c=1$)
\begin{equation}
\mathcal{L}_{\text{PQED}}=\frac{F_{\mu\nu}F^{\mu\nu}}{2(-\square)^{1/2}}+\bar{\psi}_{a}(i\gamma^{0}\partial_{0}+iv_{F}\mathbf{\gamma}\cdot\nabla)\psi_{a}+j^{\mu}A_{\mu}-\frac{\xi}{2}A_{\mu}\frac{\partial^{\mu}\partial^{\nu}}{(-\square)^{1/2}}A_{\nu},
\label{eq:dens-lag}
\end{equation}
which was first proposed by Marino \cite{marino1993quantum}, where $\square$ is the d'Alembertian operator, $F_{\mu\nu}$ is the usual electromagnetic tensor,
$v_{F}$ is the bare Fermi velocity of the electrons in graphene,
$\psi_{a}^{\dagger}=(\psi_{A\uparrow}^{*}\psi_{A\downarrow}^{*}\psi_{B\uparrow}^{*}\psi_{B\downarrow}^{*})_{a}$ is the four-component Dirac spinor representation of the electrons in the $A$ and $B$ sub-lattices of graphene, 
$\gamma^{\mu}=(\gamma^{0},v_F \mathbf{\gamma})$ are rank-$4$ Dirac matrices, 
$a$ is the flavor index which represents
the sum over $K$ and $K^{\prime}$ in the Brillouin zone, 
$j^{\mu}$
is the matter current in $2+1$ dimensions, and the last term is the gauge
fixing. 
The PQED has been successfully used in the description of several graphene properties \cite{alves2013chiral,marino2015interaction,nascimento2015chiral,menezes2016fermi,menezes2016influence,alves2017dynamical,menezes2017spin,nascimento2017introduction,silva2017inhibition,marino2018screening,marino2018quantum,alves2018two,pires2018cavity,marino2020graphene,magalhaes2020pseudo}.

Considering Eq. (\ref{eq:dens-lag}), the photon propagator is \cite{alves2013chiral}
\begin{equation}
\Delta_{\mu\nu}^{(0)}(k)=\frac{2\pi}{\kappa\sqrt{k^{2}}}\left[\delta_{\mu\nu}-\left(1-\frac{1}{\xi}\right)\frac{k_{\mu}k_{\nu}}{k^{2}}\right],
\end{equation}
where $\kappa$ is the dielectric constant of the environment in cgs units (for conversion from SI to cgs see \cite{zhang1998electromagnetic}), and we defined the quadri-momentum  $k=(k_0,\mathbf{k})$, with $\cancel{k}=\gamma^{0}k_{0}+v_{F}\mathbf{\gamma}\cdot\mathbf{k}$ and $k^{2}=k_{0}^{2}+v_{F}^{2}|\mathbf{k}|^{2}$. Also, since we are working in the Euclidean space representation, the $\gamma$-matrices can be set to satisfy $\left\{ \gamma^{\mu},\gamma^{\nu}\right\} =2\delta^{\mu\nu}\mathbf{I}$, where $\mu,\nu=0,1,2$ and $\mathbf{I}={\rm diag}(1,1,1)$  \cite{barnes2014effective}.

Assuming the Feynman gauge ($\xi=1$) and the non-retarded regime ($k_{0}=0$), we get
\begin{equation}
\Delta_{\mu\nu}^{(0)}(|\mathbf{k}|)=\frac{2\pi}{\kappa|\mathbf{k}|}\delta_{0\mu}\delta_{0\nu}, \label{gra-prop}
\end{equation}
which, by a Fourier transform \cite{roberts1994dyson}, leads to the Coulombian potential for static charges,
\begin{equation}
V(|\mathbf{r}|)=\frac{e}{\kappa|\mathbf{r}|},
\label{coulomb}
\end{equation}
where $e$ is the bare coupling constant and $|\mathbf{r}|$ is the distance between two electrons in graphene. 

\begin{figure}[t]
\centering{}\includegraphics[width=4cm]{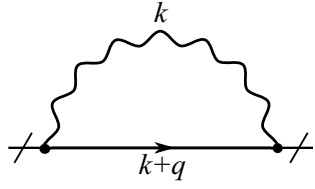}\caption{Electron self-energy diagram.
\label{fig:Diagrama-de-auto-energia-2}}
\end{figure}

Using (\ref{gra-prop}), the electron self-energy (represented by the diagram in Fig. \ref{fig:Diagrama-de-auto-energia-2}) is given by \cite{gonzalez1994non}
\begin{equation}
\Sigma(\mathbf{q})=-\frac{e^{2}}{4\kappa}\mathbf{q}\cdot\mathbf{\gamma}\ln\left(\frac{\Lambda}{|\mathbf{q}|}\right),\label{eq:self-energy-free}
\end{equation}
where $\Lambda$ is the ultraviolet momentum cutoff, which is inversely proportional to the lattice parameter of graphene \cite{barnes2014effective}. From Eq. (\ref{eq:self-energy-free}), one obtains the renormalized Fermi velocity \cite{gonzalez1994non}
\begin{equation}
v^{*}_{\mathbf{q}}=v_{F}\left[1+\frac{\alpha}{4}\ln\left(\frac{\Lambda}{|\mathbf{q}|}\right)\right],
\label{vf-q}
\end{equation}
where $\alpha={e^{2}}/({\kappa v_{F}})$ is the fine structure constant of graphene. Consequently, the renormalized $\alpha$ will be \cite{barnes2014effective}
\begin{equation}
\alpha^{*}_{\mathbf{q}}=\frac{\alpha}{1+\frac{\alpha}{4}\ln(\Lambda/|\mathbf{q}|)}.
\label{alpha_q}
\end{equation}

\begin{figure}[h]
	\centering{}\includegraphics[width=8cm]{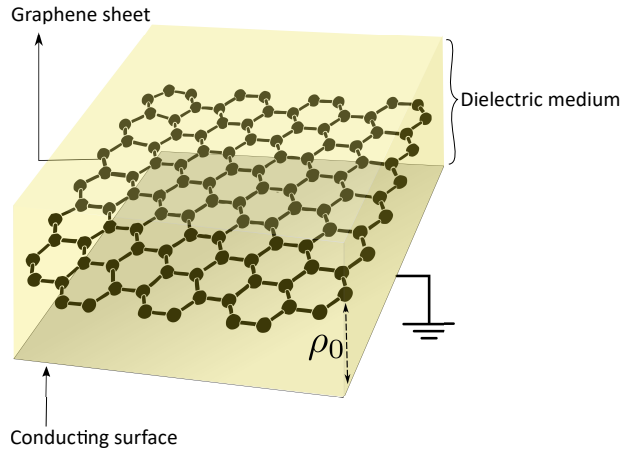}
	\caption{A graphene sheet a distance $\rho_{0}$ from a grounded perfectly conducting surface. The dielectric media above and below graphene have the same dielectric constant $\kappa$. For $\kappa=1$, one recovers the situation described in Ref. \cite{silva2017inhibition}.
	\label{fig:Graphene-sheet}}
\end{figure}

Among the various ways to influence the transport properties of graphene (see \cite{neto2009electronic,de2012space}), it has been shown by Silva \emph{et al.} \cite{silva2017inhibition,pires2018cavity} that a grounded perfectly conducting surface can inhibit the renormalization of the Fermi velocity in a neutral graphene sheet placed in vacuum (see Fig. \ref{fig:Graphene-sheet}).
In this context, we can also highlight the work by Raoux
\emph{et al.} \cite{raoux2010velocity} who obtained, for doped graphene,
the inhibition of the renormalized Fermi velocity in the same configuration.

\begin{figure}[h]
	\centering{}\includegraphics[width=7cm]{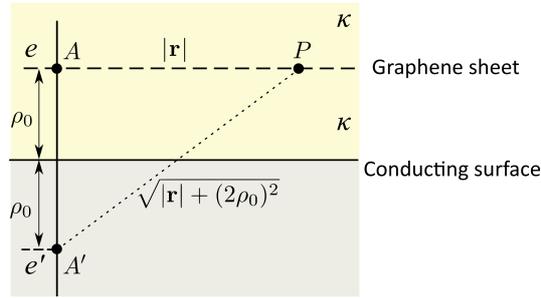}\caption{
	Illustration of the graphene sheet, represented by the dashed line, located at a distance $\rho_0$ from the conducting surface (represented by the solid horizontal line).
	An electron $e$ in graphene, located at the point $A$, has its image $e^{\prime}=-e$ at the point $A^{\prime}$. $P$ is an arbitrary point in the plane of graphene.
	\label{fig:electrons}}
\end{figure}

Considering Fig. \ref{fig:electrons}, an electron (charge $e$) in graphene (located at point $P$), in the presence of a grounded perfectly conducting plate, distant $\rho_0$ and parallel to the graphene sheet, interacts not only with another electron in the same sheet (at point $A$), but also with a certain amount of positive charge on the surface of the conducting plate induced by the other electron (at point $A^{\prime }$).  Using the image method, this amount of positive charge is effectively given by an image charge $e^{\prime} = -e$ (see Fig. \ref{fig:electrons}). Taking this into account, the effective static potential related to an electron in graphene, in the presence of a conducting plate, is given by \cite{silva2017inhibition,pires2018cavity}
\begin{equation}
V(\rho_{0},|\mathbf{r}|) =\frac{e}{\kappa}\left[\frac{1}{|\mathbf{r}|}-\frac{1}{\sqrt{|\mathbf{r}|+(2\rho_{0})^{2}}}\right].\label{eq:potential}
\end{equation}

Taking a Fourier transform over Eq. (\ref{eq:potential}), the photon propagator becomes \cite{silva2017inhibition,pires2018cavity}
\begin{equation}
\Delta_{00}^{(0)}(\rho_{0},|\mathbf{k}|)=\frac{2\pi}{\kappa|\mathbf{k}|}\left(1-e^{-2\rho_{0}|\mathbf{k}|}\right).
\label{eq:photon-prop}
\end{equation}
The Fermion propagator remains unchanged in the presence of the conducting surface, namely
\begin{equation}
S_{F}^{(0)}(k)=\frac{\gamma^{0}k_{0}+v_{F}\mathbf{\gamma}\cdot\mathbf{k}}{k_{0}^{2}+v_{F}^{2}|\mathbf{k}|^{2}}.\label{eq:electron-prop}
\end{equation}

Considering Eq. (\ref{eq:photon-prop}), the electron self-energy
correction becomes \cite{silva2017inhibition}
\begin{equation}
\Sigma(\rho_{0},\mathbf{q})=-\frac{e^{2}}{4\kappa}\mathbf{q}\cdot\mathbf{\gamma}\left[\ln\left(\frac{\Lambda}{|\mathbf{q}|}\right)-F(\rho_{0}|\mathbf{q}|,\Lambda)\right],\label{eq:self-energy}
\end{equation}
where
\begin{equation}
F(\rho_{0}|\mathbf{q}|,\Lambda)=\frac{1}{2\pi}\int_{0}^{2\pi}d\zeta\int_{0}^{\xi_{\Lambda}}d\xi(1+\cosh\xi\cos\zeta)\exp[-\rho_{0}|\mathbf{q}|(\cosh\xi-\cos\zeta)], \label{F-int}
\end{equation}
and
\begin{equation}
\xi_{\Lambda}=\cosh^{-1}\left(\frac{2\Lambda}{|\mathbf{q}|}+\cos\zeta\right).
\end{equation}
Hence, the renormalized Fermi velocity will be written as \cite{silva2017inhibition}
\begin{equation}
v^{*}_{\rho_{0},\mathbf{q}}=v_{F}\left\{ 1+\frac{\alpha}{4}\left[\ln\left(\frac{\Lambda}{|\mathbf{q}|}\right) -F(\rho_{0}|\mathbf{q}|,\Lambda)\right]\right\}.
\label{vf-rho-0-q}
\end{equation}

In Ref. \cite{pires2018cavity}, the authors considered the real part of the optical conductivity $\sigma_{{\rm opt}}(\omega)$ as presented in Ref. \cite{stauber2017interacting}:
\begin{equation}
\frac{\sigma^{*}_{{\rm opt}}(\omega)}{\sigma_{0}}=1+C \alpha^{*}_{\omega/v_{F}} \equiv 1+C \alpha\left.\frac{v_{F}}{v^{*}_{\mathbf{q}}}\right|_{|\mathbf{q}|=\omega/v_F},
\label{eq:optical}
\end{equation}
where $C$ is a constant, $\sigma_0=e^2/4$ is the minimal conductivity of graphene and the superscript in $\sigma^{*}$ sets the dependence on the renormalized parameter $\alpha^{*}_{\omega/v_{F}}$.
By replacing $v^{*}_{\mathbf{q}}$ [Eq. (\ref{vf-q})] by $v^{*}_{\rho_0,\mathbf{q}}$ [Eq. (\ref{vf-rho-0-q})] in the above equation, they concluded that the inhibition of the renormalized Fermi velocity, caused by the conducting plate, leads to an increase of the optical conductivity in graphene. 

In the present paper, we investigate this issue computing, until $2$-loop order of perturbation, the longitudinal conductivity of a neutral graphene sheet in a dielectric medium (instead of vacuum as considered in Ref. \cite{silva2017inhibition,pires2018cavity}) and near a grounded perfectly conducting surface (see Fig. \ref{fig:Graphene-sheet}). We also calculate, analytically, the conductivity in the optical limit.

The paper is organized as follows.
In Sec. \ref{sec-statement}, we state the problem, presenting the calculations to be done. Namely, we use the Kubo formula to obtain the longitudinal conductivity in terms of the $00$-component of the polarization tensor \cite{mahan2013many}.
In Sec. \ref{sec-Pi-2a}, we calculate the first of the $2$-loop diagrams that compose the $00$-component of the polarization tensor.
In Sec. \ref{sec-Pi-1-plus-Pi-2a-renor}, we rewrite the polarization function in terms of renormalized parameters, such that comparison to experimental results can be feasible.
In Sec. \ref{sec-Pi-2b}, we compute the second of the $2$-loop diagrams that composes the $00$-component of the polarization tensor.
In Sec. \ref{sec-conductivity}, we merge all results and get the longitudinal conductivity.
In Sec. \ref{sec:Optical-limit}, the optical limit of the conductivity is taken.
In Sec. \ref{sec-final}, we analyze our results and make final comments.
%

%%%%%%%%%%%%%%%%%%%%%%%%%%%%%%%%%%%%%%%%%%%%%%%%%%%%%%%%%%%%%%%%%%%%%%%%%%%%%%%%%%%%%
\section{Statement of the problem}
\label{sec-statement}

\begin{figure}[h]
\subfloat[$\Pi_{1}$ diagram.
\label{fig-Pi-1}]
{\includegraphics[width=5cm]{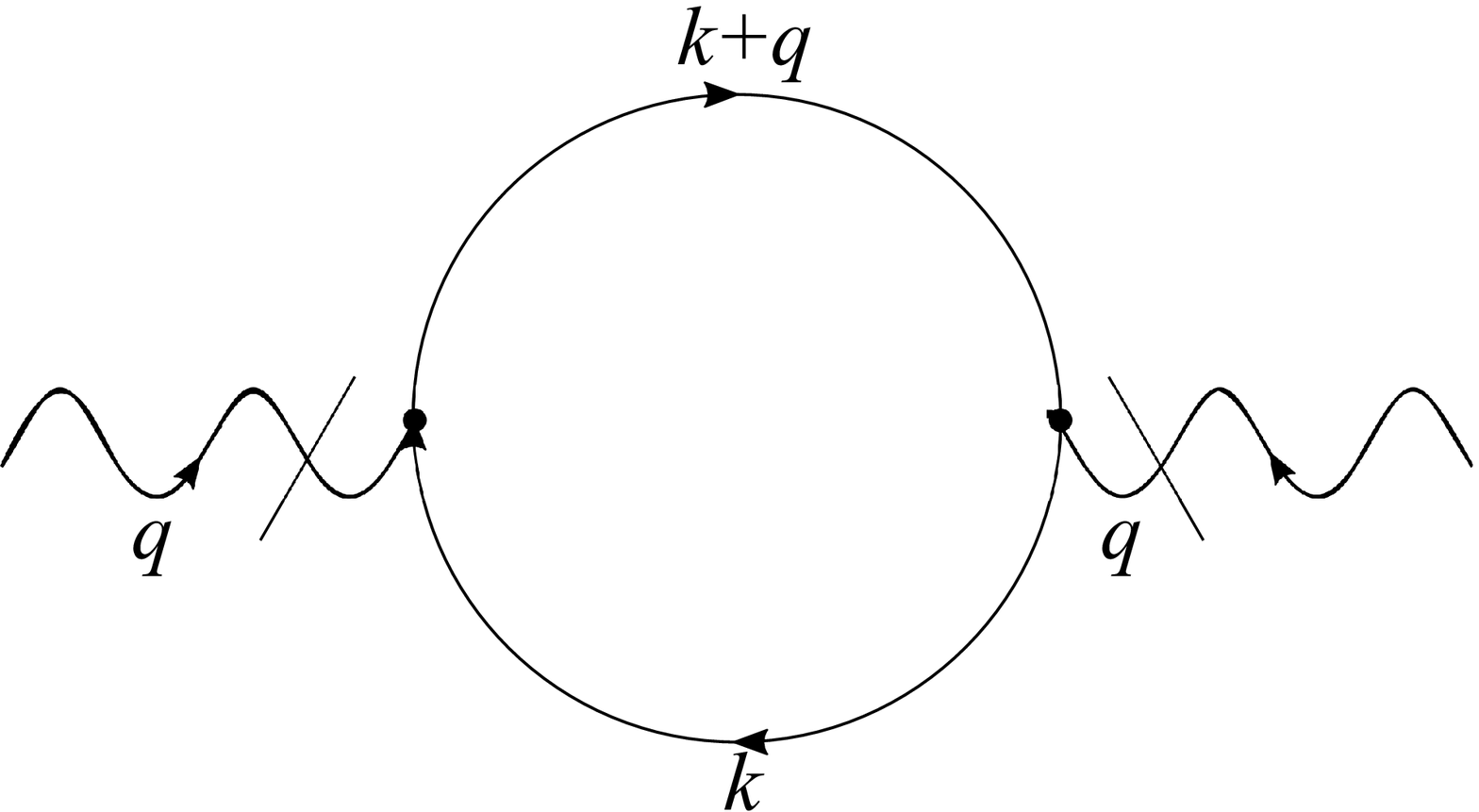}}
\hfill{}
\subfloat[$\Pi_{2a}$ diagram. 
\label{fig-Pi-2a}]
{\includegraphics[width=5cm]{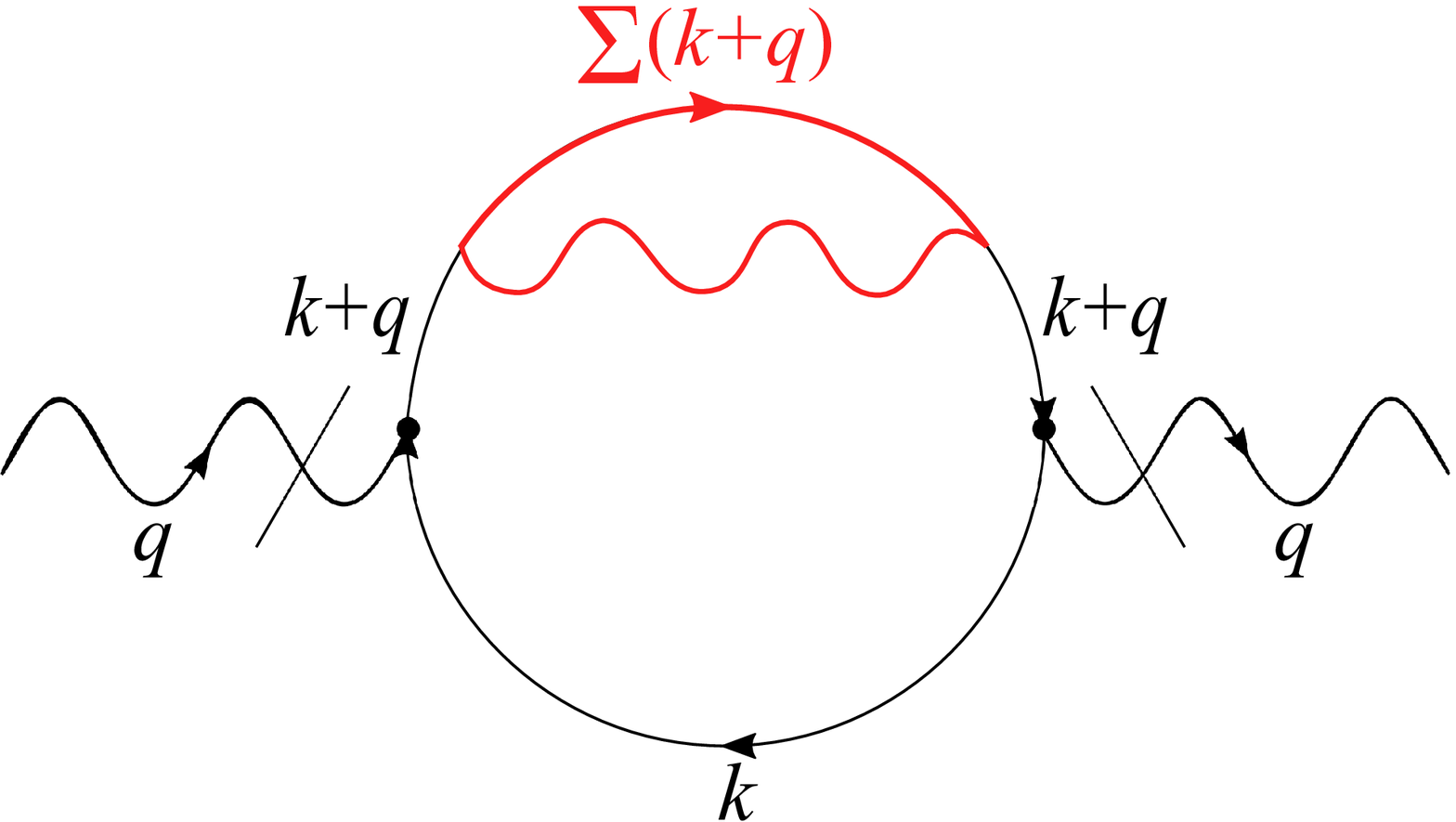}}
\hfill{}\subfloat[$\Pi_{2b}$ diagram.
\label{fig-Pi-2b}]
{\includegraphics[width=5cm]{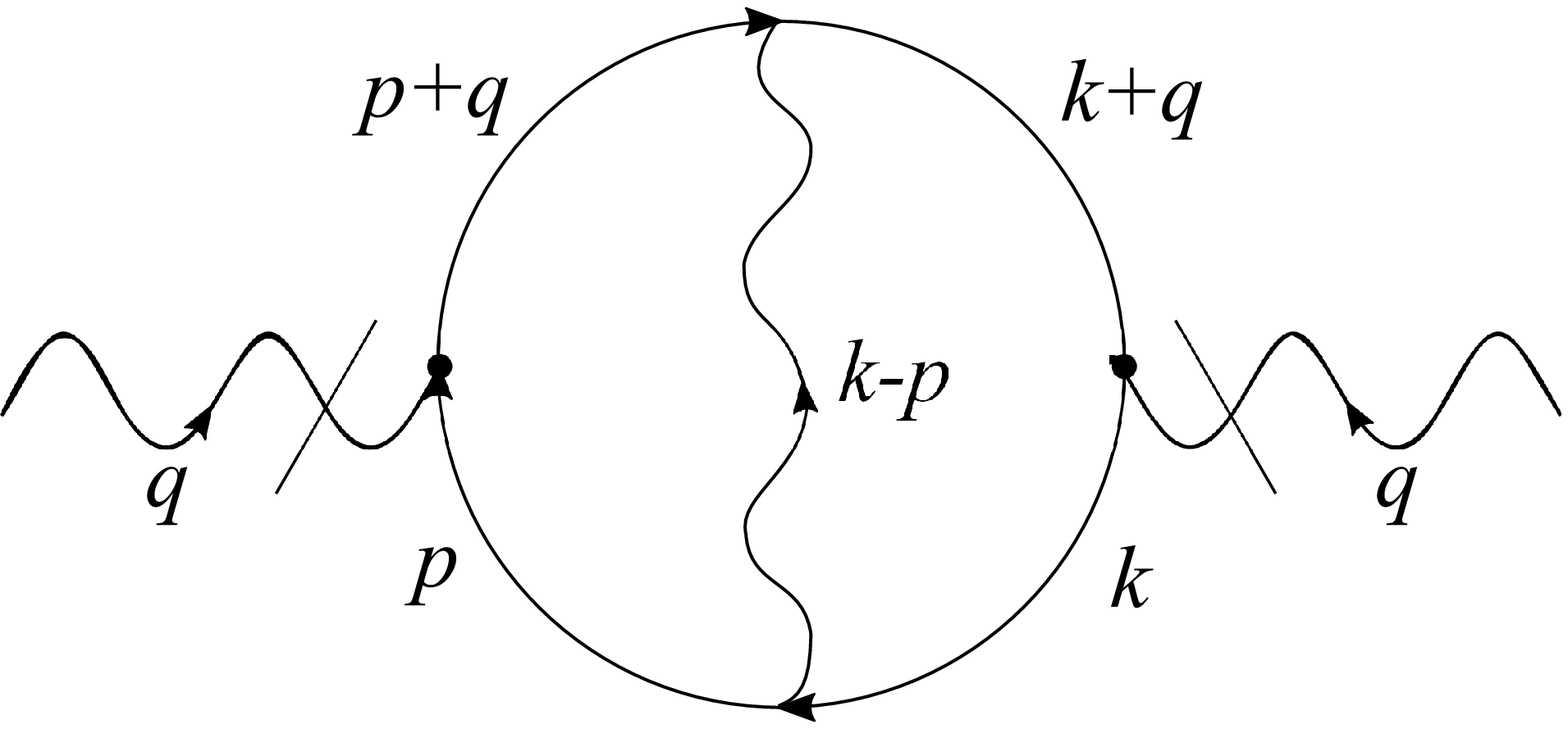}}
\caption{Feynman diagrams representing the first terms of the perturbative expansion of 
$\Pi$, according to Eq. (\ref{Pi-approx}). The red part of the diagram in Fig. \ref{fig-Pi-2a} is to highlight the contribution of the electron-self energy diagram.}
\end{figure}

We consider a neutral graphene sheet, in a dielectric medium, near a grounded perfectly conducting surface, arranged as in Fig. \ref{fig:Graphene-sheet}.
From the Kubo formula, the longitudinal conductivity, $\sigma(\omega,\mathbf{q})$, can be obtained in terms of the $00$-component  of the polarization tensor, $\Pi(\omega,\mathbf{q})$, as follows \cite{mahan2013many,sodemann2012interaction,kotov2008electron}:
\begin{equation}
\sigma(\omega,\mathbf{q})=\frac{i\omega\Pi(\omega,\mathbf{q})}{|\mathbf{q}|^{2}}.
\label{formula-kubo}
\end{equation}
The function $\Pi(\omega,\mathbf{q})$ can be expanded perturbatively as
\begin{equation}
\Pi(\omega,\mathbf{q})\approx \Pi_{1}(\omega,\mathbf{q}) + 2 \Pi_{2a}(\omega,\mathbf{q}) + \Pi_{2b}(\omega,\mathbf{q}),
\label{Pi-approx}
\end{equation}
where $\Pi_{1}$ is the contribution coming from the $1$-loop Feynman diagram (see Fig. \ref{fig-Pi-1}), whereas $\Pi_{2a}$ (Fig. \ref{fig-Pi-2a}) and $\Pi_{2b}$ (Fig. \ref{fig-Pi-2a}) are contributions from the $2$-loop diagrams. 
Due to its symmetry, the correction $\Pi_{2a}$ must be multiplied by the factor $2$. 

From Fig. \ref{fig-Pi-1}, one can see that there are no photon lines in the $\Pi_{1}$ diagram, therefore, the result is the same for QED in $2+1$ dimensions, given by \cite{pisarski1984chiral,appelquist1985chiral,barnes2014effective}

\begin{equation}
\Pi_{1}(\omega,\mathbf{q})=-\frac{Ne^{2}|\mathbf{q}|}{8v_{F}}\frac{1}{\sqrt{1-y^{2}_{q}}},
\label{eq:Pi-B-1}
\end{equation}
where $N=2$ corresponds to the $\mathcal{K}$ and $\mathcal{K^{\prime}}$ valleys, and we defined ${iq_{0}}=\omega+i\epsilon$,  $\epsilon\rightarrow0^{+}$, with $y_{q}=(\omega+i\epsilon)/v_{F}|\mathbf{q}|$  (remember that $q=(q_0,\mathbf{q})$ represents the quadri-momentum).

The diagrams in Figs. \ref{fig-Pi-2a} and \ref{fig-Pi-2b} show that these contributions to $\Pi$ depend on the photon propagator and, therefore, are directly affected by the presence of the conducting plate, which is a distance $\rho_0$ of the graphene sheet [to emphasize this influence, hereafter, we write $\Pi_{2a}(\rho_{0},\omega,\mathbf{q})$ and $\Pi_{2b}(\rho_{0},\omega,\mathbf{q})$].

In the present paper, the main calculation is to obtain the $2$-loop contributions to $\Pi$,
represented by the diagrams shown in Figs. \ref{fig-Pi-2a} 
and \ref{fig-Pi-2b}, in the presence of a grounded perfectly conducting plate as shown in Fig. \ref{fig:Graphene-sheet}.
Then, the longitudinal conductivity, according to the Kubo formula (\ref{formula-kubo}), can be obtained.

%%%%%%%%%%%%%%%%%%%%%%%%%%%%%%%%%%%%%%%%%%%%%%%%%%%%%%%%%%%%%%%%%%%%%%%%%%%%%%%%%%%
\section{Calculation of $\Pi_{2a}$}
\label{sec-Pi-2a}

The diagram presented in Fig. \ref{fig-Pi-2a} leads to the following
definition:
\begin{align}
\Pi_{2a}(\rho_{0},\omega,\mathbf{q}) & =-N\int\frac{d^{3}k}{(2\pi)^{3}}{\rm Tr}\left[e\gamma^{0}S_{F}^{(0)}(k)e\gamma^{0}S_{F}^{(0)}(k+q)\Sigma(\rho_{0},\mathbf{k}+\mathbf{q})S_{F}^{(0)}(k+q)\right],
\end{align}
which, from Eqs. (\ref{eq:photon-prop}), (\ref{eq:electron-prop}) and (\ref{eq:self-energy}), gives
\begin{align}
\Pi_{2a}(\rho_{0},\omega,\mathbf{q}) & =Ne^{2}\int\frac{d^{3}k}{(2\pi)^{3}}{\rm Tr}\left\{ \gamma^{0}\frac{\cancel{k}}{k^{2}}\gamma^{0}\frac{(\cancel{k}+\cancel{q})}{(k+q)^{2}}\frac{e^{2}}{4\kappa}(\mathbf{k}+\mathbf{q})\cdot\mathbf{\gamma}\left[\ln\left(\frac{\Lambda}{|\mathbf{k}+\mathbf{q}|}\right)-F(\rho_{0}|\mathbf{k}+\mathbf{q}|,\Lambda)\right]\right.\nonumber \\
 & \times\left.\frac{(\cancel{k}+\cancel{q})}{(k+q)^{2}}\right\} .\label{Pi-AE-1}
\end{align}
Following the same approach of Ref. \cite{barnes2014effective},
one can obtain that
\begin{align}
\Pi_{2a}(\rho_{0},\omega,\mathbf{q}) & =-\frac{Ne^{4}}{2\kappa}\int\frac{d^{2}k}{\left(2\pi\right)^{2}}\frac{\mathbf{k}\cdot(\mathbf{k}+\mathbf{q})-|\mathbf{k}||\mathbf{k}+\mathbf{q}|}{|\mathbf{k}|}\frac{\left[v_{F}^{2}\left(|\mathbf{k}|+|\mathbf{k}+\mathbf{q}|\right)^{2}-q_{0}^{2}\right]}{\left[v_{F}^{2}\left(|\mathbf{k}|+|\mathbf{k}+\mathbf{q}|\right)^{2}+q_{0}^{2}\right]^{2}}\nonumber \\
 & \times\left[\ln\left(\frac{\Lambda}{|\mathbf{k}+\mathbf{q}|}\right)-F(\rho_{0}|\mathbf{k}+\mathbf{q}|,\Lambda)\right].\label{eq:Pi-2a-cartesian}
\end{align}
By making $\mathbf{k}\rightarrow-\mathbf{k}-\mathbf{q}$,
choosing a coordinate system such that $\mathbf{q}=(|\mathbf{q}|,0)$,
and performing a transformation to elliptic coordinates \cite{barnes2014effective,sodemann2012interaction},
\begin{equation}
k_{x}=\frac{|\mathbf{q}|}{2}(\cosh\mu\cos\nu-1),\qquad k_{y}
=
\frac{|\mathbf{q}|}{2}\sinh\mu\sin\nu,\qquad d^{2}k=\frac{|\mathbf{q}|^{2}}{4}(\cosh^{2}\mu-\cos^{2}\nu)d\mu d\nu,
\label{eq:elliptic-coord}
\end{equation}
we are able to obtain
\begin{align}
\Pi_{2a}(\rho_{0},\omega,\mathbf{q}) & =\frac{Ne^{4}|\mathbf{q}|}{32\pi^{2}\kappa v_{F}^{2}}\int_{0}^{2\pi}d\nu\int_{0}^{\infty}d\mu\frac{\sin^{2}\nu(\cosh\mu-\cos\nu)(\cosh^{2}\mu+y^{2}_{q})}{(\cosh^{2}\mu-y^{2}_{q})^{2}}\nonumber \\
 & \times\left[\ln\left(\frac{2\Lambda}{|\mathbf{q}|\left(\cosh\mu-\cos\nu\right)}\right)-F\left(\frac{\rho_{0}|\mathbf{q}|}{2}(\cosh\mu-\cos\nu),\Lambda\right)\right].\label{PI-AE-2}
\end{align}
The above equation can be rewritten as
\begin{equation}
\Pi_{2a}(\rho_{0},\omega,\mathbf{q})=\frac{e^{2}\alpha|\mathbf{q}|}{16\pi v_{F}}\left[\frac{\pi}{2}\frac{1}{(1-y^{2}_{q})^{3/2}}\ln(\Lambda/|\mathbf{q}|)+I_{a^{\prime}}(y_{q})-I_{a^{\prime\prime}}(\rho_{0}|\mathbf{q}|,y_{q},\Lambda)\right],\label{eq:Pi-AE}
\end{equation}
where the function $I_{a^{\prime}}(y_{q})$ was, to our best knowledge, first obtained in Ref. \cite{sodemann2012interaction}:
\begin{align}
I_{a^{\prime}}(y_q) & =\frac{1}{3}\frac{1+2y^{2}_{q}}{1-y^{2}_{q}}-\frac{y_{q}}{6}\frac{5-2y^{2}_{q}}{1-y^{2}_{q}}\ln\left(\frac{1-y_{q}}{1+y_{q}}\right)-\frac{\pi}{12}\frac{3-12\ln2+6y^{2}_{q}-4y^{4}_{q}}{(1-y^{2}_{q})^{3/2}}-\frac{i}{(1-y^{2}_{q})^{3/2}}\nonumber \\
 & \times\left[\frac{\pi^{2}}{4}-{\rm Li}_{2}\left(y_{q}+i\sqrt{1-y^{2}_{q}}\right)+{\rm Li}_{2}\left(-y_{q}-i\sqrt{1-y^{2}_{q}}\right)+\frac{i\pi}{2}\ln\left(y_{q}+i\sqrt{1-y^{2}_{q}}\right)\right],
\end{align}
with ${\rm Li}_{2}(z)=\sum_{k=1}^{\infty}z^{k}/k^{2}$ being the dilogarithmic function, whereas $I_{a^{\prime\prime}}(\rho_{0}|\mathbf{q}|,y_q,\Lambda)$ is the following integral, which will be solved numerically:
\begin{equation}
I_{a^{\prime\prime}}(\rho_{0}|\mathbf{q}|,y_{q},\Lambda)=\int_{0}^{2\pi}d\nu\int_{0}^{\infty}d\mu\frac{\sin^{2}\nu(\cosh\mu-\cos\nu)(\cosh^{2}\mu+y^{2}_{q})}{\pi(\cosh^{2}\mu-y^{2}_{q})^{2}}F\left(\frac{\rho_{0}|\mathbf{q}|}{2}(\cosh\mu-\cos\nu),\Lambda\right).\label{eq:Ib-pi2a}
\end{equation}
Defining $x_q={\rm Re}\,y_{q}=\omega/(v_{F}|\mathbf{q}|)$, we see that, for $x_q>1$, there is a pole to contour. In such situation, we can determine the real and imaginary parts of $I_{a^{\prime\prime}}(\rho_{0}|\mathbf{q}|,y_q,\Lambda)$ using a generalization of the Sokhotski-Plemelj identity for higher order poles \cite{galapon2016cauchy}, as presented in \ref{subsec:Pi2a-diagram}:
\begin{equation}
{\rm Re}\,I_{a^{\prime\prime}}(\rho_{0}|\mathbf{q}|,x_q,\Lambda) =\lim_{\epsilon\rightarrow 0^{+}}\int_{0}^{2\pi}d\nu\left[\int_{1}^{x_q-\epsilon}\frac{H(w,x_q,\nu)}{(w-x_q)^{2}}dw+\int_{x_q+\epsilon}^{\infty}\frac{H(w,x_q,\nu)}{(w-x_q)^{2}}dw-\frac{2H(x_q,x_q,\nu)}{\epsilon}\right],
\end{equation}
\begin{equation}
{\rm Im}\,I_{a^{\prime\prime}}(\rho_{0}|\mathbf{q}|,x_q,\Lambda) =\int_{0}^{2\pi}d\nu\left.\frac{dH(w,x_q,\nu)}{dw}\right|_{w=x_q},
\end{equation}
where $w=\cosh\mu$ and
\begin{equation}
H(w,x_q,\nu) =\frac{1}{\pi}\frac{\sin^{2}\nu(w-\cos\nu)(w^{2}+x^{2}_{q})}{(w+x_q)^{2}\sqrt{w^{2}-1}}F\left(\frac{\rho_{0}|\mathbf{q}|}{2}(w-\cos\nu),\Lambda\right).
\end{equation}

%%%%%%%%%%%%%%%%%%%%%%%%%%%%%%%%%%%%%%%%%%%%%%%%%%%%%%%%%%%%%%%%%%%%%%%%%%%%%%%%%%%
\section{The representation of $\Pi_{1}+2\Pi_{2a}$ in terms of renormalized parameters} \label{sec-Pi-1-plus-Pi-2a-renor}

The $\Pi_{1}$ and $\Pi_{2a}$ functions are written in terms of bare parameters and the momentum cutoff $\Lambda$, which, as argued in Ref. \cite{sodemann2012interaction}, can be arbitrarily chosen. On the other hand, the  renormalized parameters are observable quantities. Therefore, to go from bare to renormalized parameters, we follow a procedure used in Refs. \cite{barnes2014effective,sodemann2012interaction,kotov2008electron}. From Eqs. (\ref{eq:Pi-B-1}) and (\ref{eq:Pi-AE}), we write
\begin{align}
\Pi_{1}(\omega,\mathbf{q})+2\Pi_{2a}(\rho_{0},\omega,\mathbf{q}) & =-\frac{Ne^{2}|\mathbf{q}|}{8v_{F}}\left\{ \frac{1}{\sqrt{1-y^{2}_{q}}}-\frac{\alpha}{4}\frac{1}{(1-y^{2}_{q})^{3/2}}\left[\ln(\Lambda/|\mathbf{q}|)-F(\rho_{0}|\mathbf{q}|,\Lambda)\right]-\frac{\alpha}{4}\frac{F(\rho_{0}|\mathbf{q}|,\Lambda)}{(1-y^{2}_{q})^{3/2}}\right.\nonumber \\
 & \left.-\frac{\alpha}{2\pi}\left[I_{a^{\prime}}(y_q)-I_{a^{\prime\prime}}(\rho_{0}|\mathbf{q}|,y_q,\Lambda)\right]\right\},
\label{Pi-1-plus-Pi-2a}
\end{align}
where, for convenience, we added and subtracted ${Ne^{2}|\mathbf{q}|}{\alpha}{F(\rho_{0}|\mathbf{q}|,\Lambda)}/[{{{32v_{F}}}(1-y^{2}_{q})^{3/2}}]$ to the right hand side.

The $F$ dependence on $\Lambda$ can removed if we consider  $\rho_0$ much greater than the lattice parameter, such that the integral in (\ref{F-int}) is equivalent to assume $\Lambda\rightarrow \infty$, as stated in Ref. \cite{silva2017inhibition}. Then, we can use the following  approximation \cite{silva2017inhibition}:
\begin{equation}
F(\rho_{0}|\boldsymbol{q}|,\Lambda)\approx I_{0}(\rho_{0}|\boldsymbol{q}|)K_{0}(\rho_{0}|\boldsymbol{q}|)+I_{1}(\rho_{0}|\boldsymbol{q}|)K_{1}(\rho_{0}|\boldsymbol{q}|),
\end{equation}
where $I_{\nu}$ and $K_{\nu}$ are the modified Bessel functions of first and second kind, respectively, and, hereafter, we remove the $\Lambda$ dependence from $F$, writing $F(\rho_{0}|\boldsymbol{q}|)$.

Next, considering the renormalized Fermi velocity in Eq. (\ref{vf-rho-0-q}) and the renormalized fine structure constant of graphene in the presence of a conducting plate, defined as
\begin{equation}
\alpha^{*}_{\rho_{0},\mathbf{q}}=\frac{\alpha}{1+\frac{\alpha}{4}[\ln(\Lambda/|\mathbf{q}|)-F(\rho_{0}|\mathbf{q}|)]},
\label{alpha_rho_q}
\end{equation}
we rewrite Eq. (\ref{Pi-1-plus-Pi-2a}) in terms of renormalized parameters by computing $\alpha$ in terms of $\alpha^{*}_{\rho_{0},\mathbf{q}}$ and applying a series expansion until $\mathcal{O}(\alpha^{*}_{\rho_{0},\mathbf{q}})$, obtaining
\begin{equation}
\Pi_{1}(\omega,\mathbf{q})+2\Pi_{2a}(\rho_{0},\omega,\mathbf{q})\approx -\frac{e^{2}|\mathbf{q}|}{4v^{*}_{\rho_{0},\mathbf{q}}\sqrt{1-y_{\rho_{0},q}^{*2}}}+\frac{e^{2}|\mathbf{q}|\alpha^{*}_{\rho_{0},\mathbf{q}}}{v^{*}_{\rho_{0},\mathbf{q}}}p_{2a}(\rho_{0}|\mathbf{q}|,y^{*}_{\rho_{0},q}),
\label{eq:pi1+pi2b}
\end{equation}
where
\begin{equation}
p_{2a}(\rho_{0}|\mathbf{q}|,y^{*}_{\rho_{0},q}) =\frac{1}{8\pi}\left[\frac{\pi}{2}\frac{F(\rho_{0}|\mathbf{q}|)}{(1-y_{\rho_{0},q}^{* 2})^{3/2}}+I_{a^{\prime}}(y^{*}_{\rho_{0},q})-I_{a^{\prime\prime}}(\rho_{0}|\mathbf{q}|,y^{*}_{\rho_{0},q})\right],\label{eq:p2a-eq}
\end{equation}
\begin{equation}
y^{*}_{\rho_{0},q}=\frac{\omega+i\epsilon}{v^{*}_{\rho_{0},\mathbf{q}}|\mathbf{q}|},\qquad\text{and}\qquad x^{*}_{\rho_{0},q}=\frac{\omega}{v^{*}_{\rho_{0},\mathbf{q}}|\mathbf{q}|}.
\end{equation}

%%%%%%%%%%%%%%%%%%%%%%%%%%%%%%%%%%%%%%%%%%%%%%%%%%%%%%%%%%%%%%%%%%%%%%%%%%%%%%%%%%%
\section{Calculation of $\Pi_{2b}$}
\label{sec-Pi-2b}

The diagram in Fig. \ref{fig-Pi-2b} leads to the following polarization correction:
\begin{equation}
\Pi_{2b}(\rho_{0},\omega,\mathbf{q}) =
-Ne^{4}\iint\frac{d^{3}k}{\left(2\pi\right)^{3}}\frac{d^{3}p}{\left(2\pi\right)^{3}}\Delta_{00}^{(0)}(\rho_{0},|\mathbf{k}-\mathbf{p}|)
{\rm Tr} \left[\gamma^{0}S_{F}^{(0)}(p+q)\gamma^{0}S_{F}^{(0)}(k+q)\gamma^{0}S_{F}^{(0)}(k)\gamma^{0}S_{F}^{(0)}(p)\right],
\end{equation}
which, according to Eqs. (\ref{eq:photon-prop}) and (\ref{eq:electron-prop}), gives
\begin{equation}
\Pi_{2b}(\rho_{0},\omega,\mathbf{q})=-\frac{2\pi e^{4}N}{\kappa}\iint\frac{d^{3}k}{\left(2\pi\right)^{3}}\frac{d^{3}p}{\left(2\pi\right)^{3}}\frac{1-e^{-2\rho_{0}|\mathbf{k}-\mathbf{p}|}}{|\mathbf{k}-\mathbf{p}|}{\rm Tr}\left[\gamma^{0}\frac{\left(\cancel{p}+\cancel{q}\right)}{\left(p+q\right)^{2}}\gamma^{0}\frac{\left(\cancel{k}+\cancel{q}\right)}{\left(k+q\right)^{2}}\gamma^{0}\frac{\cancel{k}}{k^{2}}\gamma^{0}\frac{\cancel{p}}{p^{2}}\right].\label{eq:PI-VE}
\end{equation}

Considering Ref. \cite{barnes2014effective}, we find
\begin{align}
\Pi_{2b}(\rho_{0},\omega,\mathbf{q}) & =-\frac{2\pi Ne^{4}}{\kappa}\iint\frac{d^{2}k}{(2\pi)^{2}}\frac{d^{2}p}{(2\pi)^{2}}\frac{1-e^{-2\rho_{0}|\mathbf{k}-\mathbf{p}|}}{|\mathbf{k}-\mathbf{p}|[v_{F}^{2}(|\mathbf{k}|+|\mathbf{k}+\mathbf{q}|)^{2}+q_{0}^{2}][v_{F}^{2}(|\mathbf{p}|+|\mathbf{p}+\mathbf{q}|)^{2}+q_{0}^{2}]}\nonumber \\
 & \times\left\{ -q_{0}^{2}\left(\frac{\mathbf{k}\cdot\mathbf{p}}{|\mathbf{k}||\mathbf{p}|}-\frac{\mathbf{p}\cdot(\mathbf{k}+\mathbf{q})}{|\mathbf{p}||\mathbf{k}+\mathbf{q}|}-\frac{\mathbf{k}\cdot(\mathbf{p}+\mathbf{q})}{|\mathbf{k}||\mathbf{p}+\mathbf{q}|}+\frac{(\mathbf{k}+\mathbf{q})\cdot(\mathbf{p}+\mathbf{q})}{|\mathbf{k}+\mathbf{q}||\mathbf{p}+\mathbf{q}|}\right)\right.\nonumber \\
 & +\frac{v_{F}^{2}(|\mathbf{k}|+|\mathbf{k}+\mathbf{q}|)(|\mathbf{p}|+|\mathbf{p}+\mathbf{q}|)}{|\mathbf{k}||\mathbf{p}||\mathbf{k}+\mathbf{q}||\mathbf{p}+\mathbf{q}|}\left[\mathbf{k}\cdot\mathbf{p}|\mathbf{q}|^{2}-(\mathbf{k}\cdot\mathbf{q})(\mathbf{p}\cdot\mathbf{q})\right.\nonumber \\
 & \left.\left.+\left(|\mathbf{k}|^{2}+\mathbf{k}\cdot\mathbf{q}-|\mathbf{k}||\mathbf{k}+\mathbf{q}|\right)\left(|\mathbf{p}|^{2}+\mathbf{p}\cdot\mathbf{q}-|\mathbf{p}||\mathbf{p}+\mathbf{q}|\right)\right]\right\} .\label{PI-V-2}
\end{align}
By making a transformation to elliptic coordinates over momenta $\mathbf{k}$ and $\mathbf{p}$, in the same way of Eq. (\ref{eq:elliptic-coord}), we get
\begin{align}
\Pi_{2b}(\rho_{0},\omega,\mathbf{q})= & -\frac{Ne^{2}|\mathbf{q}|\alpha}{16\pi^{3}v_{F}}\int\frac{d\mu d\mu^{\prime}d\nu d\nu^{\prime}\cosh\mu\cosh\mu^{\prime}\sin\nu\sin\nu^{\prime}}{\sqrt{[\cosh(\mu+\mu^{\prime})-\cos(\nu+\nu^{\prime})][\cosh(\mu-\mu^{\prime})-\cos(\nu-\nu^{\prime})]}}\nonumber \\
\times & \frac{\left[\sin\nu\sin\nu^{\prime}+\sinh\mu\sinh\mu^{\prime}+y_{q}^{2}\left(\sin\nu\sin\nu^{\prime}+\tanh\mu\tanh\mu^{\prime}\cos\nu\cos\nu^{\prime}\right)\right]}{\left(\cosh^{2}\mu-y_{q}^{2}\right)\left(\cosh^{2}\mu^{\prime}-y_{q}^{2}\right)}\nonumber \\
\times & \left[1-\exp\left(-\rho_{0}|\mathbf{q}|\sqrt{[\cosh(\mu+\mu^{\prime})-\cos(\nu+\nu^{\prime})][\cosh(\mu-\mu^{\prime})-\cos(\nu-\nu^{\prime})]}\right)\right].
\label{eq:PI-V-3}
\end{align}
After that, making $\eta=\nu+\nu^{\prime}$, $\tau=\nu-\nu^{\prime}$, $a=\mu+\mu^{\prime}$ and $b=\mu-\mu^{\prime}$, we find:
\begin{align}
\Pi_{2b}(\rho_{0},\omega,\mathbf{q}) & =\frac{|\mathbf{q}|e^{2}\alpha}{v_{F}}p_{2b}(\rho_{0}|\mathbf{q}|,y_q),\label{eq:Pi2bfunc}
\end{align}
where
\begin{align}
 p_{2b}(\rho_{0}|\mathbf{q}|,y_q) & =-\frac{1}{16\pi^{3}}\int_{0}^{\infty}db\int_{b}^{\infty}da\frac{1}{\left[1-2y_{q}^{2}+\cosh\left(a+b\right)\right]\left[1-2y_{q}^{2}+\cosh\left(a-b\right)\right]}\nonumber \\
 & \times\left[(\cosh^{2}a-\cosh^{2}b)(1+y_{q}^{2})I_{1}(a,b,\rho_0|\mathbf{q}|)+(\cosh a+\cosh b)I_{2}(a,b,\rho_0|\mathbf{q}|)\right.\nonumber \\
 &
 \left.+y_{q}^{2}(\cosh a-\cosh b)I_{3}(a,b,\rho_0|\mathbf{q}|)\right],\label{eq:p2b}
\end{align}
and
\begin{equation}
I_{i}(a,b,\rho_0|\mathbf{q}|)=\int_{0}^{2\pi}d\eta\int_{0}^{2\pi}d\tau\,h_{i}\,(1-h_{\rho_0|\mathbf{q}|}),\quad\text{for}\quad i=1,2,3,
\end{equation}
with
\begin{equation}
h_{1}=\frac{\cos\tau-\cos\eta}{\sqrt{[\cosh a-\cos\eta][\cosh b-\cos\tau]}},\qquad h_{2}=\frac{\left(\cos\tau-\cos\eta\right)^{2}}{\sqrt{[\cosh a-\cos\eta][\cosh b-\cos\tau]}},
\end{equation}
\begin{equation}
h_{3}=\frac{\cos^{2}\tau-\cos^{2}\eta}{\sqrt{[\cosh a-\cos\eta][\cosh b-\cos\tau]}},\qquad h_{\rho_0|\mathbf{q}|}=\exp\left(-\rho_{0}|\mathbf{q}|\sqrt{[\cosh a-\cos\eta][\cosh b-\cos\tau]}\right).
\end{equation}

The integral in Eq. (\ref{eq:p2b}) has a pole at $x_q>1$. Therefore, within this region, we split the integral into real and imaginary parts with the application of the Sokhotski-Plemelj identity (as shown in \ref{subsec:Pi2b-diagram}), resulting in
\begin{align}
&{\rm Re}[p_{2b}(\rho_{0}|\mathbf{q}|,x_q)]  =-\frac{1}{16\pi^{3}}\lim_{\epsilon\rightarrow 0^{+}}\int_{0}^{\infty}db\left[\int_{b}^{\lambda(x_q)-b-\epsilon}da\frac{U(a,b,\rho_0|\mathbf{q}|,\lambda(x_q))}{g_{+}(a,b,\lambda(x_q))g_{-}(a,b,\lambda(x_q))}\right.\nonumber \\
 & +\left.\int_{\lambda(x_q)-b+\epsilon}^{\lambda(x_q)+b-\epsilon}da\frac{U(a,b,\rho_0|\mathbf{q}|,\lambda(x_q))}{g_{+}(a,b,\lambda(x_q))g_{-}(a,b,\lambda(x_q))}+\int_{\lambda(x_q)+b+\epsilon}^{\infty}da\frac{U(a,b,\rho_0|\mathbf{q}|,\lambda(x_q))}{g_{+}(a,b,\lambda(x_q))g_{-}(a,b,\lambda(x_q))}\right],
\end{align}
\begin{align}
{\rm Im}[p_{2b}(\rho_{0}|\mathbf{q}|,x_q)]& =-\frac{1}{32\pi^{2}\sinh(\lambda(x_q))}\left\{ \int_{0}^{\lambda(x_q)/2}\frac{db}{\sinh b}\left[\frac{U(\lambda(x_q)+b,b,\rho_0|\mathbf{q}|,\lambda(x_q))}{\sinh(b+\lambda(x_q))}\right.\right.\nonumber \\
&
\left.\left.-\frac{U(\lambda(x_q)-b,b,\rho_0|\mathbf{q}|,\lambda(x_q))}{\sinh(\lambda(x_q)-b)}\right] +\int_{\lambda(x_q)/2}^{\infty}\frac{db}{\sinh b}\frac{U(\lambda(x_q)+b,b,\rho_0|\mathbf{q}|,\lambda(x_q))}{\sinh(b+\lambda(x_q))}\right\} ,
\end{align}
where, following Ref. \cite{sodemann2012interaction}, we have defined $x_q=\cosh(\lambda/2)$, and
\begin{align}
U(a,b,\rho_0|\mathbf{q}|,\lambda)& =(\cosh^{2}a-\cosh^{2}b)I_{1}(a,b,\rho_0|\mathbf{q}|)+\left(\frac{\cosh\lambda+3}{2}\right)(\cosh a +\cosh b)I_{2}(a,b,\rho_0|\mathbf{q}|) \nonumber \\
&
 +\frac{\cosh\lambda+1}{2}(\cosh a-\cosh b)I_{3}(a,b,\rho_0|\mathbf{q}|).
\end{align}
with
\begin{equation}
g_{+}(a,b,\lambda)=\cosh(a+b)-\cosh\lambda,
\end{equation}
and 
\begin{equation}
g_{-}(a,b,\lambda)=\cosh(a-b)-\cosh\lambda.
\end{equation}

Finally, we consider the map $(\alpha,v_{F})\rightarrow(\alpha^{*}_{\rho_{0},\mathbf{q}},v^{*}_{\rho_{0},\mathbf{q}})$ in Eq. (\ref{eq:Pi2bfunc}), rewriting $\Pi_{2b}$ in terms of these renormalized parameters until $\mathcal{O}(\alpha^{*}_{\rho_{0},\mathbf{q}})$. Then, we have
\begin{equation}
\Pi_{2b}(\rho_{0},\omega,\mathbf{q}) =\frac{e^{2}|\mathbf{q}|\alpha^{*}_{\rho_{0},\mathbf{q}}}{v^{*}_{\rho_{0},\mathbf{q}}}p_{2b}(\rho_{0}|\mathbf{q}|,y^{*}_{\rho_0,q}).
\label{eq-Pi-2b}
\end{equation}

%%%%%%%%%%%%%%%%%%%%%%%%%%%%%%%%%%%%%%%%%%%%%%%%%%%%%%%%%%%%%%%%%%%%%%%%%%%%%%%%%%
\section{The conductivity}
\label{sec-conductivity}

By inserting Eqs. (\ref{eq:pi1+pi2b}) and (\ref{eq-Pi-2b}) into (\ref{Pi-approx}), we can write the $00$-component of the polarization tensor, corrected until $2$-loop order, in terms of the renormalized parameters, namely
\begin{equation}
\Pi(\rho_{0},\omega,\mathbf{q})\approx-\frac{e^{2}|\mathbf{q}|}{4 v^{*}_{\rho_{0},\mathbf{q}}\sqrt{1-y_{\rho_{0},q}^{* 2}}}+\frac{e^{2}|\mathbf{q}|\alpha^{*}_{\rho_{0},\mathbf{q}}}{v^{*}_{\rho_{0},\mathbf{q}}}[p_{2a}(\rho_{0}|\mathbf{q}|,y^{*}_{\rho_{0},q})+p_{2b}(\rho_{0}|\mathbf{q}|,y^{*}_{\rho_{0},q})].\label{eq:Pi-total}
\end{equation}
From Eqs. (\ref{formula-kubo}) and (\ref{eq:Pi-total}), the longitudinal conductivity becomes
\begin{equation}
\frac{\sigma(\rho_{0},\omega,\mathbf{q})}{\sigma_{0}} \approx 4i x^{*}_{\rho_{0},q}\left\{ -\frac{1}{4\sqrt{1-y_{\rho_{0},q}^{* 2}}}+\alpha^{*}_{\rho_{0},\mathbf{q}}[p_{2a}(\rho_{0}|\mathbf{q}|,y^{*}_{\rho_{0},q})+p_{2b}(\rho_{0}|\mathbf{q}|,y^{*}_{\rho_{0},q})]\right\}.
\label{eq-conductivity-ratio}
\end{equation}
In order to better visualize the results, it will be useful to rewrite $x^{*}_{\rho_{0},q}$ in terms of $x^{*}_{q}$, which is done using Eqs. (\ref{vf-q}) and (\ref{vf-rho-0-q}), remembering that $\alpha^{*}_{\mathbf{q}}=e^2/(\kappa v^{*}_\mathbf{q})$, giving
\begin{equation}
	x_{\rho_{0},q}^{*} = \frac{x_{q}^{*}}{\left[1-\frac{\alpha^{*}_{\mathbf{q}}}{4}F(\rho_{0}\mathbf{q})\right]}.
	\label{x_rho}
\end{equation}
Also, we can rewrite $\alpha^{*}_{\rho_{0},\mathbf{q}}$ in terms of $\alpha^{*}_{\mathbf{q}}$, using Eqs. (\ref{alpha_q}) and (\ref{alpha_rho_q}) we get
\begin{equation}
\alpha^{*}_{\rho_{0},\mathbf{q}}=\frac{\alpha^{*}_{\mathbf{q}}}{1-\frac{\alpha^{*}_{\mathbf{q}}}{4}F(\rho_{0}|\mathbf{q}|)}.
\label{alpha_rho_alpha_q}	
\end{equation}

For further analysis, taking into account the real part of the longitudinal conductivity of graphene, it is convenient to define
\begin{equation}
\tilde{\sigma}_{1}(\rho_0|\mathbf{q}|,x_{q}^{*})=-{\rm Re}\left\{ \frac{i x_{q}^{*}}{\sqrt{\left[1-\frac{\alpha^{*}_{\mathbf{q}}}{4}F(\rho_{0}\mathbf{q})\right]^2-y_{q}^{*2}}}\right\},
\label{eq-sigma-tilde-1}
\end{equation}
which is the real part of the ratio (\ref{eq-conductivity-ratio}), calculated until 1-loop order of perturbation (the subscript 1 means that the formula takes into account calculations up to 1-loop). We also define:
\begin{align}
\tilde{\sigma}_2(\rho_0|\mathbf{q}|,x_{q}^{*})= &{\rm Re}\left\{ -\frac{ix_{q}^{*}}{\sqrt{\left[1-\frac{\alpha^{*}_{\mathbf{q}}}{4}F(\rho_{0}|\mathbf{q}|)\right]^{2}-y_{q}^{*2}}}+\frac{4ix_{q}^{*}\alpha_{\mathbf{q}}^{*}}{\left[1-\frac{\alpha^{*}_{\mathbf{q}}}{4}F(\rho_{0}\mathbf{q})\right]^{2}}\left[p_{2a}\left(\rho_{0}|\mathbf{q}|,\frac{y_{q}^{*}}{1-\frac{\alpha^{*}_{\mathbf{q}}}{4}F(\rho_{0}\mathbf{q})}\right)\right.\right.\nonumber \\
& \left.\left.+p_{2b}\left(\rho_{0}|\mathbf{q}|,\frac{y_{q}^{*}}{1-\frac{\alpha^{*}_{\mathbf{q}}}{4}F(\rho_{0}\mathbf{q})}\right)\right]\right\}.
\label{eq-sigma-tilde}
\end{align}
which is the real part of the ratio (\ref{eq-conductivity-ratio}), written in terms of $x_{q}^{*}$ and $\alpha^{*}_{\mathbf{q}}$ using Eqs. (\ref{x_rho}) and (\ref{alpha_rho_alpha_q}). The subscript $2$ means we are going until $2$-loop perturbation order.

\begin{figure}[h]
	\subfloat[
	\label{fig:Re-cond}]
	{\includegraphics[width=8cm]{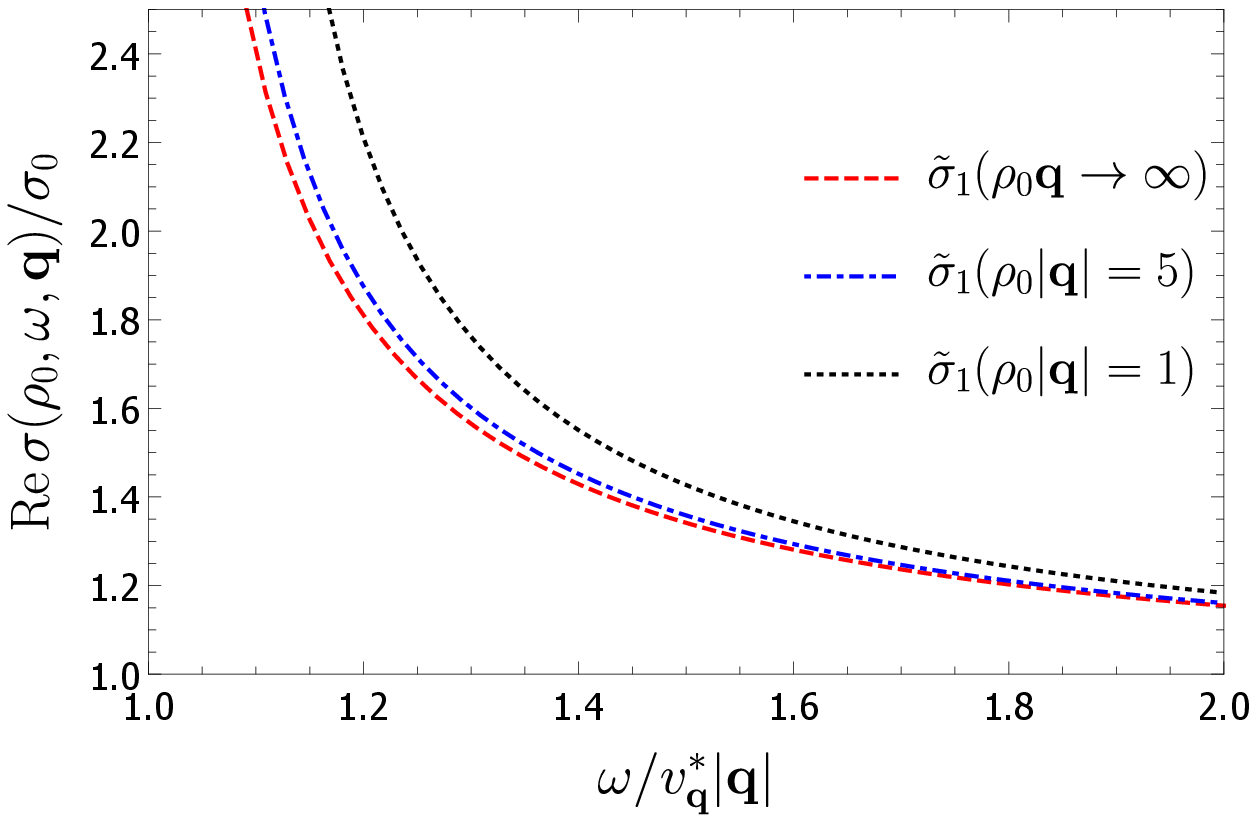}}
	\hfill{}
	\subfloat[
	\label{fig:Re-cond2}]
	{\includegraphics[width=8cm]{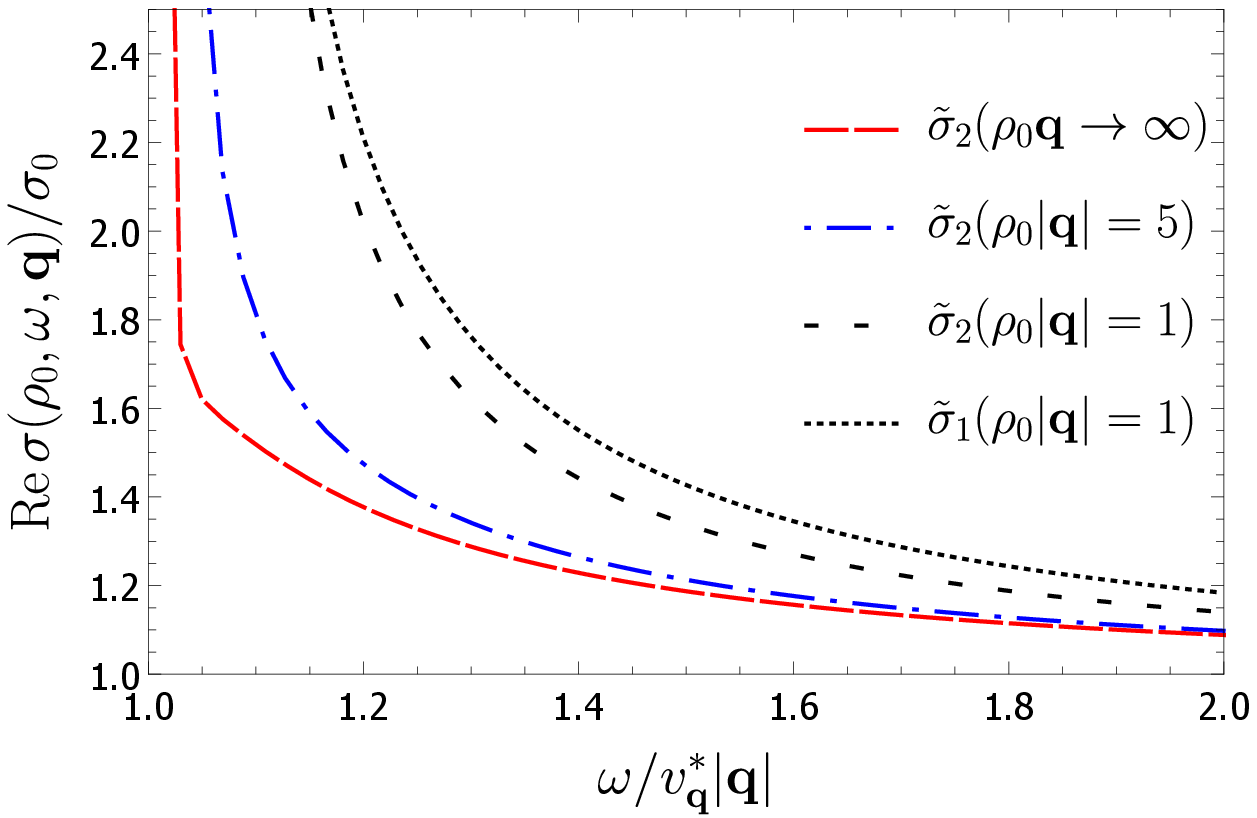}}
	\caption{Panel \ref{fig:Re-cond}, representation of the real part of the conductivity at $1$-loop perturbation order, where the dashed line accounts for $\tilde{\sigma}_{1}(\rho_0|\mathbf{q}|\rightarrow \infty,x_{q}^{*})$, the dot-dashed line for $\tilde{\sigma}_{1}(\rho_0|\mathbf{q}|=5,x_{q}^{*})$ and the dotted line for $\tilde{\sigma}_{1}(\rho_0|\mathbf{q}|=1,x_{q}^{*})$. Panel \ref{fig:Re-cond2} presents the curves for the longitudinal conductivity up to $2$-loop perturbation order, where the long-dashed line accounts for $\tilde{\sigma}_{2}(\rho_0|\mathbf{q}|\rightarrow \infty,x_{q}^{*})$, the dot-long-dashed line for $\tilde{\sigma}_{2}(\rho_0|\mathbf{q}|=5,x_{q}^{*})$ and the small dashed line for $\tilde{\sigma}_{2}(\rho_0|\mathbf{q}|=1,x_{q}^{*})$, we also plotted $\tilde{\sigma}_{1}(\rho_0|\mathbf{q}|=1,x_{q}^{*})$ for comparison purposes. Both panels have $\omega/(v^{*}_{\mathbf{q}}|\mathbf{q}|)$ as the horizontal axis. We made $\alpha^{*}_{\mathbf{q}}=0.3$ as in Ref. \cite{sodemann2012interaction}.
	}
	
\end{figure}

In Fig. \ref{fig:Re-cond}, we plotted the real part of the conductivity at $1$-loop order, represented by $\tilde{\sigma}_{1}(\rho_0|\mathbf{q}|,x_{q}^{*})$, given in Eq. (\ref{eq-sigma-tilde-1}). The three curves correspond to $\tilde{\sigma}_{1}(\rho_0|\mathbf{q}|\rightarrow \infty,x_{q}^{*})$ (dashed line), $\tilde{\sigma}_{1}(\rho_0|\mathbf{q}|=5,x_{q}^{*})$ (dot-dashed line) and $\tilde{\sigma}_{1}(\rho_0|\mathbf{q}|=1,x_{q}^{*})$ (dotted line). From this panel we can take two conclusions: first, as we bring the conducting plate closer to the graphene sheet, the $1$-loop conductivity increases, with such effect being more noticeable as we approach the threshold ($\omega/v^{*}_{\mathbf{q}}|\mathbf{q}|$);
% This behavior is directly correlated to the inhibition of the renormalization of the Fermi velocity, which implies that $x_{q}^{*}<x_{\rho_0,q}^{*}$, as can be seen from Eq. (\ref{x_rho}).
%
second, for a fixed $\rho_0$, we also see an increase in the conductivity as $|\mathbf{q}|$ decreases, which is a consequence of the product $\rho_0|\mathbf{q}|$ in the exponential term of the photon propagator (\ref{eq:photon-prop}).

In Fig. \ref{fig:Re-cond2}, we plotted the real part of the conductivity up to $2$-loop perturbation, represented by $\tilde{\sigma}_{2}(\rho_0|\mathbf{q}|,x_{q}^{*})$, given by Eq. (\ref{eq-sigma-tilde}), and, for comparison, the $1$-loop correction for $\rho_0|\mathbf{q}|=1$, i.e.  $\tilde{\sigma}_{1}(\rho_0|\mathbf{q}|=1,x_{q}^{*})$.
The $2$-loop curves correspond to $\tilde{\sigma}_{2}(\rho_0|\mathbf{q}|\rightarrow \infty,x_{q}^{*})$ (long dashed line), $\tilde{\sigma}_{2}(\rho_0|\mathbf{q}|=5,x_{q}^{*})$ (dot-long-dashed  line) and $\tilde{\sigma}_{2}(\rho_0|\mathbf{q}|=1,x_{q}^{*})$ (small dashed line).
When the product $\rho_0|\mathbf{q}|$ decreases, we observe an inhibition of the $2$-loop correction, which, therefore, causes a displacement of the longitudinal conductivity towards the $1$-loop correction, leading to an increase in the conductivity even more evident than at $1$-loop.
In other words, as the distance between the conducting plate and graphene gets smaller, the contribution of the $2$-loop polarization diagrams becomes inhibited, and this leads to an enhancement of graphene's conductivity for this part of the spectrum (in the optical limit we see a slightly different behavior).
On the other hand, for a fixed $\rho_0$, if we vary $|\mathbf{q}|$ and $\omega$ such that the ratio $\omega/(v^{*}_{\mathbf{q}}|\mathbf{q}|)$ remains constant, the previous behavior will also be observed: a cancellation of the $2$-loop corrections as $|\mathbf{q}|$ decreases, leading to an increase of $\tilde{\sigma}_{2}(\rho_{0}|\mathbf{q}|,x_q)$.
We must highlight that this feature is not observed in graphene in the absence of the plate, since the only dependence it has on the external momentum comes from $\omega/(v^{*}_{\mathbf{q}}|\mathbf{q}|)$, but varying $|\mathbf{q}|$ while keeping this quantity constant will not affect the conductivity.

We also remark that, according to Ref. \cite{sodemann2012interaction}, as we get closer to the threshold $\omega/(v^{*}_{\mathbf{q}}|\mathbf{q}|)=1$, less the perturbation up to 2-loop order is in good agreement with experimental results, requiring, for this, the consideration of higher order terms in the perturbative expansion.
%

%%%%%%%%%%%%%%%%%%%%%%%%%%%%%%%%%%%%%%%%%%%%%%%%%%%%%%%%%%%%%%%%%%%%%%%%%%%%%%%%%%%%%%%%%%%%%%%%%%%%%
\section{The optical limit of the conductivity}
\label{sec:Optical-limit}

The optical conductivity, $\sigma_{{\rm opt}}$, is defined by \cite{mishchenko2008minimal}
\begin{equation}
\sigma_{{\rm opt}}(\rho_0,\omega)= \lim_{|\mathbf{q}|\rightarrow 0}\frac{i\omega}{|\mathbf{q}|^{2}}\Pi(\rho_0,\omega,\mathbf{q}),
\label{formula-kubo-opt}
\end{equation}
which can be written, using Eq. (\ref{Pi-approx}), as
\begin{equation}
\sigma_{{\rm opt}}(\rho_0,\omega)\approx \sigma_{{\rm opt},1}(\omega) + \sigma_{{\rm opt},2a}(\rho_0,\omega) 
+ \sigma_{{\rm opt},2b}(\rho_0,\omega),
\label{eq-sigma-opt-approx}
\end{equation}
where
\begin{equation}
\sigma_{{\rm opt},1}(\omega) = \lim_{|\mathbf{q}|\rightarrow 0}\frac{i\omega}{|\mathbf{q}|^{2}}\Pi_{1}(\omega,\mathbf{q}),
\label{eq-sigma-opt-1}
\end{equation}
\begin{equation}
\sigma_{{\rm opt},2a}(\rho_0,\omega) = \lim_{|\mathbf{q}|\rightarrow 0}\frac{i\omega}{|\mathbf{q}|^{2}}2\Pi_{2a}(\rho_0,\omega,\mathbf{q}),
\label{eq-sigma-opt-2a}
\end{equation}
\begin{equation}
\sigma_{{\rm opt},2b}(\rho_0,\omega) = \lim_{|\mathbf{q}|\rightarrow 0}\frac{i\omega}{|\mathbf{q}|^{2}}\Pi_{2b}(\rho_0,\omega,\mathbf{q}).
\label{eq-sigma-opt-2b}
\end{equation}

From Eq. (\ref{eq:Pi-B-1}), we get that $\sigma_{{\rm opt},1}(\omega)$ will be given by
\begin{equation}
\sigma_{{\rm opt},1}(\omega)=\lim_{|\mathbf{q}|\rightarrow 0}\frac{i\omega}{|\mathbf{q}|^{2}}\Pi_{1}(\omega,\mathbf{q})=\frac{e^{2}}{4}=\sigma_{0},
\label{eq-sigma-opt-1-sol}
\end{equation}
so that the contribution of $\sigma_{{\rm opt},1}$ results in the minimal conductivity $\sigma_0$.
The real parts of the $2$-loop contributions to the conductivity, calculated in \ref{sec:opt-limit}, lead to
 \begin{equation}
 {\rm Re}[\sigma_{{\rm opt},2a}(\rho_{0},\omega)]=\sigma_{0}\frac{\alpha}{4}\left(1+\frac{\rho_{0}\omega}{v_{F}}\frac{1}{2}\left.\frac{d F(\rho_{0}|\mathbf{k}|)}{d(\rho_{0}|\mathbf{k}|)}\right|_{|\mathbf{k}|=\omega/2v_{F}}\right),
 \label{eq-sigma-opt-2a-sol-1} 
 \end{equation}
and
\begin{equation}
 {\rm Re}[\sigma_{{\rm opt},2b}(\rho_0,\omega)]=\sigma_{0}\alpha\left[\frac{8-3\pi}{6}+\int_{0}^{\pi}\frac{d\theta}{\pi}\int_{0}^{\infty}du\frac{\cos\theta(u+\cos\theta)\exp\left(-\frac{\rho_{0}\omega}{v_{F}}\sqrt{u^{2}+1-2u\cos\theta}\right)}{\left(1-u^{2}\right)\sqrt{u^{2}+1-2u\cos\theta}}\right],
\label{eq-sigma-opt-2b-sol-1}
\end{equation}
where $u=2v_F |\mathbf{k}|/\omega$.
Hence, as in the rest of the paper, making $\rho_0 \rightarrow \infty$ recovers the results of the literature \cite{mishchenko2008minimal}:

\begin{equation}
	 {\rm Re}[\sigma_{{\rm opt},2a}(\rho_{0}\rightarrow\infty,\omega)]=\frac{1}{4}\sigma_{0}\alpha \qquad \text{and} \qquad {\rm Re}[\sigma_{{\rm opt},2b}(\rho_0\rightarrow\infty,\omega)]=\frac{8-3\pi}{6}\sigma_{0}\alpha.
\end{equation}

By replacing Eqs. (\ref{eq-sigma-opt-1-sol}), (\ref{eq-sigma-opt-2a-sol-1}) and (\ref{eq-sigma-opt-2b-sol-1}) in (\ref{eq-sigma-opt-approx}), we get 
\begin{equation}
\tilde{\sigma}_{{\rm opt}}(\rho_{0},\omega)=\frac{{\rm Re}[\sigma_{{\rm opt}}(\rho_{0},\omega)]}{\sigma_{0}}\approx 1+C(\rho_{0}\omega/v_F)\alpha,
\label{eq-sigma-tilde-opt-approx-sol}
\end{equation}
where
\begin{equation}
C(\rho_{0}\omega/v_F)  =C_{0}+\frac{\rho_{0}\omega}{8 v_{F}}\left.\frac{d}{d(\rho_{0}|\mathbf{k}|)}F(\rho_{0}|\mathbf{k}|)\right|_{|\mathbf{k}|=\omega/2v_{F}}+ \int_{0}^{\pi}\frac{d\theta}{\pi}\int_{0}^{\infty}du\frac{\cos\theta(u+\cos\theta)\exp\left(-\frac{\rho_{0}\omega}{v_{F}}\sqrt{u^{2}+1-2u\cos\theta}\right)}{\left(1-u^{2}\right)\sqrt{u^{2}+1-2u\cos\theta}},
 \label{eq-C}
\end{equation}
and
\begin{equation}
C_{0}=\frac{19-6\pi}{12}\approx 0.0125
\end{equation}
is the $2$-loop term calculated in Ref. \cite{mishchenko2008minimal}.

Considering the limit $\rho_0\rightarrow\infty$, we get
\begin{equation}
\lim_{\rho_0\rightarrow\infty}C(\rho_{0}\omega/v_F)=C_0,
\label{eq-limit-C}
\end{equation}
and
\begin{equation}
\tilde{\sigma}_{{\rm opt}}(\infty,\omega)=\lim_{\rho_0\rightarrow\infty}\tilde{\sigma}_{{\rm opt}}(\rho_{0},\omega) = 1+C_0\alpha.
\label{eq-sigma-0-tilde-opt-approx-sol}
\end{equation}

Indeed, comparing the new result $\tilde{\sigma}_{{\rm opt}}(\rho_0,\omega)$ of the present paper [taking into account the presence of a conducting plate, as shown in Eq. (\ref{eq-sigma-tilde-opt-approx-sol})] with the result from the literature, $\tilde{\sigma}_{{\rm opt}}(\infty,\omega)$ [without a conducting plate, shown in Eq. (\ref{eq-sigma-0-tilde-opt-approx-sol})], one can see that, in the former case, $C$ is frequency-dependent, whereas in the latter, it is not. Such dependence is reminiscent from the momentum dependence of the exponential term in the propagator, Eq. (\ref{eq:photon-prop}), created by the presence of the conducting plate, which, in the optical limit, was replaced by $\omega/v_F$, as show in \ref{sec:opt-limit}.

Considering the limit $\rho_0\rightarrow 0$, we have
\begin{equation}
\lim_{\rho_0\rightarrow 0}C(\rho_{0}\omega/v_F)=0,
\label{eq-limit-C-rho-0-zero}
\end{equation}
\begin{equation}
\lim_{\rho_0\rightarrow 0}\tilde{\sigma}_{{\rm opt}}(\rho_{0},\omega)= \tilde{\sigma}_{{\rm opt},1}(\omega)/\sigma_0=1.
\label{eq-sigma-0-tilde-opt-approx-sol-rho-0-zero}
\end{equation}
From this limit, we observe that there is a cancellation of the $2$-loop corrections as the distance between graphene and the conducting plate becomes negligible.

\begin{figure}[t]
	\subfloat[ Optical conductivity
	\label{fig:cond-opt}]
	{\includegraphics[width=8cm]{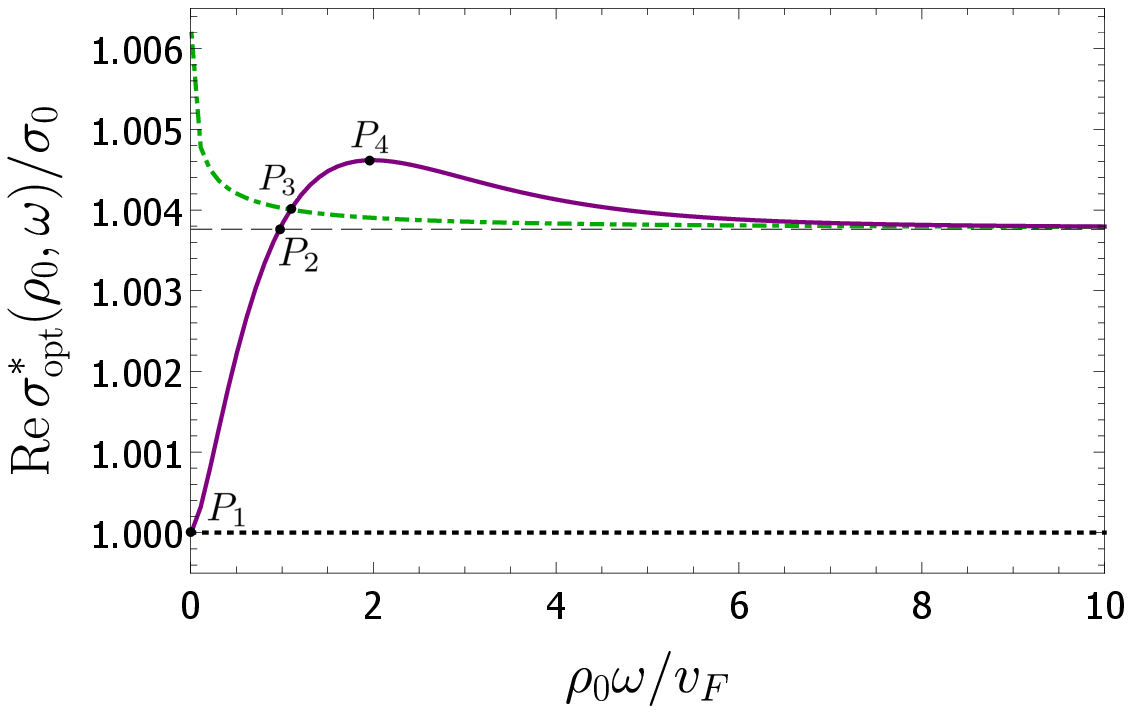}}
	\hfill{}
	\subfloat[$C$ factor
	\label{fig:crho}]
	{\includegraphics[width=8cm]{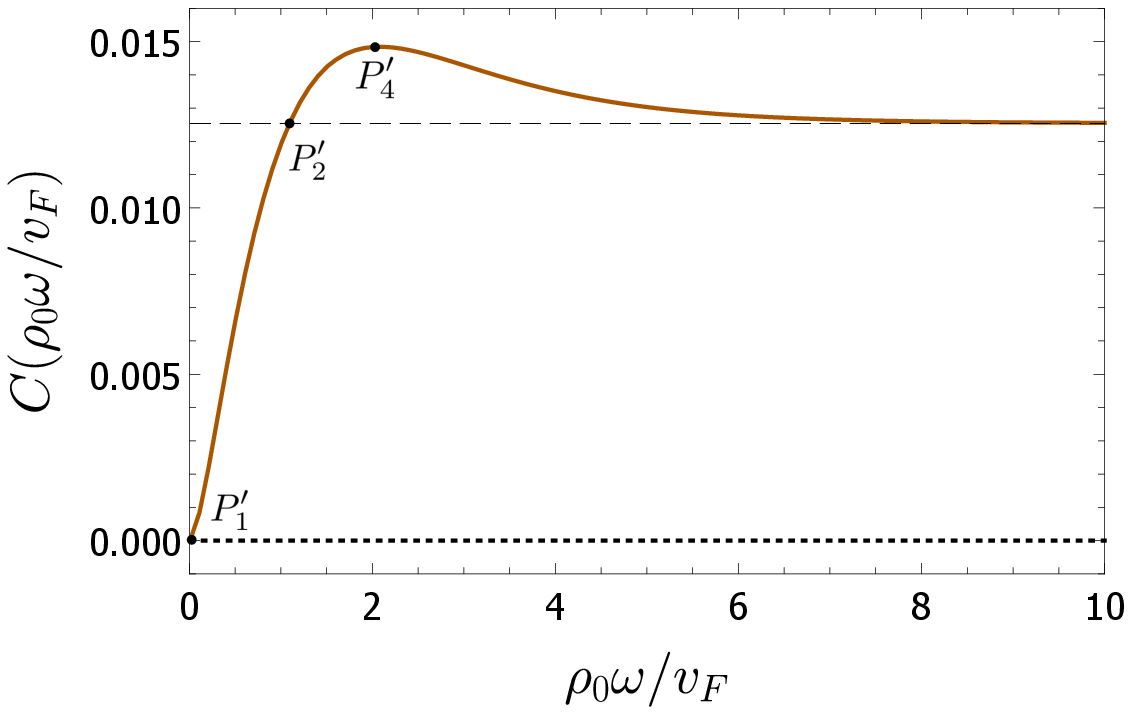}}
\caption{
In Fig. \ref{fig:cond-opt}, considering $\alpha^{*}_{\omega/v_{F}}=0.3$, the solid line corresponds to the real part of the optical conductivity, given by Eq. (\ref{eq-opt-renorm}), whereas
the dot-dashed line represents Eq. (\ref{eq:optical}), considering that $C=C_0$ and replacing $v_\mathbf{q}^{*}\rightarrow v_{\rho_0,\mathbf{q}}^{*}$. 
The dashed line serves as a reference for the value of the optical conductivity of graphene until $2$-loop order, 
without the presence of the plate.
The dotted line is a reference for the value corresponding to the minimal conductivity, represented by
the point $P_1$.
$P_2$ is the point where the values, with or without plate, for the optical conductivity are the same.
The point $P_3$ is where our result coincide with the conjecture raised in Ref. \cite{pires2018cavity}.
$P_4$ represents the maximum value of the optical conductivity.
In Fig. \ref{fig:crho}, the solid line corresponds to $C(\rho_0 \omega/v_F)$,
given by Eq. (\ref{eq-C}). 
The dashed line represents the value $C_0\approx 0.01$, 
and the dotted line serves as a reference for $C(\rho_0 \omega/v_F)\approx 0$.
At the point $P_{4}^{\prime}$, $C(\rho_0 \omega/v_F)$ reaches a maximum value, while at the point $P_{2}^{\prime}$ we have 
$C(\rho_0 \omega/v_F)=C_0 $. 
At the point $P_{1}^{\prime}$, $C(\rho_0 \omega/v_F)$ becomes null.
}
\end{figure}

By comparing (\ref{eq-sigma-0-tilde-opt-approx-sol}) with (\ref{eq:optical}), we see that the former equation is computed in terms of bare parameters and the latter is given in terms of renormalized ones. In the context of graphene with no plate \cite{mishchenko2008minimal,gonzalez1994non}, one makes $\alpha \rightarrow \alpha^{*}_{\omega/(2v_F)}$, therefore, considering the existence of the plate, we must make $\alpha \rightarrow \alpha^{*}_{\rho_0,\omega/(2v_F)}$, obtaining:
\begin{equation}
\frac{{\rm Re}[\sigma^{*}_{{\rm opt}}(\rho_0,\omega)]}{\sigma_{0}}=1+C(\rho_{0}\omega/v_{F})\alpha_{\rho_{0},\omega/v_{F}}^{*},
\label{eq-opt-renorm}
\end{equation}
which is correspondent to Eq. (\ref{eq:optical}).
Therefore, the above equation is the proper description of the optical conductivity of graphene in the presence of a conducting surface, and not just considering $v_{\mathbf{q}}^{*} \rightarrow v_{\rho_0,\mathbf{q}}^{*}$ in Eq. (\ref{eq:optical}), as supposed in Ref. \cite{pires2018cavity}.

In Fig. \ref{fig:cond-opt}, we plot the real part of the optical conductivity as a function of $\rho_0\omega/v_F$. The solid line corresponds to Eq. (\ref{eq-opt-renorm}), where the $C$ factor is dependent on the distance $\rho_0$ and the frequency $\omega$ [$C=C(\rho_0\omega/v_F)$].
In this figure, the point $P_1$ is where the optical conductivity is equal to the minimal conductivity, which means a total cancellation of the $2$-loop correction due to the presence of the plate. The point $P_2$ is where the optical conductivity in the presence of the plate equals the optical conductivity without the plate.
The point $P_4$ represents the peak of conductivity.

For comparison purposes, we also show the conductivity based on the conjecture raised in Ref. \cite{pires2018cavity}
(see the dot-dashed line in Fig. \ref{fig:cond-opt}), which considers in Eq. (\ref{eq:optical}) that the $C$ factor is a constant ($C=C_0\approx 0.01$), and the fine structure constant is given by $\alpha^{*}_{\rho_{0},\omega/v_{F}}$ defined in Eq. (\ref{alpha_rho_q}).
This approach leads to an increase of the optical conductivity of graphene as the distance $\rho_0$ decreases.
Although this is qualitatively correct for $\rho_0 \omega/v_F$ in the region 
$(P_{2x},\infty)$ [we are considering $P_i=(P_{ix},P_{iy})$ in Fig. \ref{fig:cond-opt}], the values shown by the dot-dashed line are below those indicated by the continuous line
in Fig. \ref{fig:cond-opt}.
In other words, the increase of the conductivity was under-estimated in this region,
whereas, for the region $(P_{3x},P_{2x})$, their assumption is over-estimated in comparison to the solid line.
In the region $(P_{4x},P_{3x})$, the conductivity described by the dot-dashed line keeps increasing as $\rho_0$ decreases. The solid line is now smaller than the case without the conducting plate (dashed line). This shows that the optical conductivity is not always greater than that found in the case without the plate.

In Fig. \ref{fig:crho}, the solid line corresponds to the function $C(\rho_0 \omega/v_F)$, given by Eq. (\ref{eq-C}).
For $\rho_0 \rightarrow \infty$, we recover the $C_0\approx 0.01$ obtained in Ref. \cite{mishchenko2008minimal}, which is represented by the dashed line.
As the distance $\rho_0$ becomes smaller, the function $C(\rho_0 \omega/v_F)$ increases until reaching a maximum value, given by the point $P_{4}^{\prime}$.
If we continue to shorten the distance $\rho_0$, then we have $C(\rho_0 \omega/v_F)$ coinciding with $C_0$ (point $P_{2}^{\prime}$).
After that, $C(\rho_0 \omega/v_F)$ becomes smaller than $C_0$, tending to zero in the ideal limit $\rho_0 \rightarrow 0$, which corresponds to the point $P_{1}^{\prime}$, and whose value is also indicated by the dotted line.

%%%%%%%%%%%%%%%%%%%%%%%%%%%%%%%%%%%%%%%%%%%%%%%%%%%%%%%%%%%%%%%%%%%%%%%%%%%%%%%%%%%%%%%%%%%%%%%%%%%
\section{Analysis of the results and final comments}\label{sec-final}

In the present paper, we obtained the longitudinal conductivity of a graphene sheet in the presence of a grounded perfectly conducting surface, computing also the optical limit.
The Kubo formula [Eq. (\ref{formula-kubo})] was the method used by us to compute the conductivity, with the polarization tensor 
calculated until $2$-loop perturbation order, using Pseudo-Quantum Electrodynamics to describe the interaction between electrons in graphene. 
This procedure furnished the longitudinal conductivity for any frequency and momentum [Eq. (\ref{eq-conductivity-ratio})], 
leading, in the limit $|\mathbf{q}|\rightarrow0$, to the optical conductivity [Eq. (\ref{formula-kubo-opt})].

One of our main results is that the real part of the longitudinal conductivity increases when the distance between the graphene sheet and the plate decreases.
In Fig. \ref{fig:Re-cond}, we see this enhancement for the $1$-loop approximation [Eq. (\ref{eq-sigma-tilde-1})],
but it is more evident if we go up to $2$-loop perturbation order [Eq. (\ref{eq-sigma-tilde})], specifically near the threshold $\omega/(v^{*}_{\rho_0,\mathbf{q}}|\mathbf{q}|)$,
as shown in Fig. \ref{fig:Re-cond2}.

In Ref. \cite{pires2018cavity}, considering Eq. (\ref{eq:optical}), it is suggested that the optical conductivity of a graphene sheet near a conducting plate is increased (if compared to the case without the plate) due to the inhibition of the renormalization of the Fermi velocity by the plate.
However, the authors didn't consider the influence of the plate in the $C$ factor of Eq. (\ref{eq:optical}). The calculations up to $2$-loop perturbation order provide the correct  $C$ for the model, given by Eq. (\ref{eq-C}) and shown in Fig. \ref{fig:crho}.
According to our results, even writing the conductivity in terms of the bare parameters (as shown in Eq. (\ref{eq-sigma-tilde-opt-approx-sol})), the presence of the conducting plate generates a dependence on the frequency, which does not happen in the situation without the plate.

Therefore, considering contributions from both the $C$ factor and the renormalization of the Fermi velocity, we obtain a proper description of the optical conductivity [Eq. (\ref{eq-opt-renorm})], presented in Fig. \ref{fig:cond-opt} as the solid line, with the dot-dashed line representing the conjecture indicated in Ref. \cite{pires2018cavity}. 
In this panel, when the plate is infinitely distant from the graphene sheet, 
we recover [see Eqs. (\ref{eq-C}) - (\ref{eq-sigma-0-tilde-opt-approx-sol})]
the result for the optical conductivity found in the literature \cite{mishchenko2008minimal},
which is indicated by the dashed line in Fig. \ref{fig:cond-opt}.
As we bring the conducting plate closer to the graphene sheet, the optical conductivity is increased (if compared to the case
without the presence of a conducting plate), reaching a maximum value, 
shown by the point $P_4$.
After that, the optical conductivity decreases, reaching the value of the case without the plate, as indicated by the point $P_2$.
As $\rho_0$ keeps decreasing, the value of the optical conductivity tends to the mininimal conductivity $\sigma_0$, indicated by the point $P_1$.

In summary, our results give a theoretical description of the longitudinal and optical conductivities of graphene in the presence of a conducting plate.
With calculations taken until $2$-loop perturbation order, we showed that the longitudinal conductivity increases as we bring a conducting surface closer to the graphene sheet. 
In the optical limit, the  conductivity can increase or decrease, depending on the position of the conducting plate.
These results may be useful as an alternative way to control the longitudinal and optical conductivities of graphene.

%In summary, our results give a theoretical description of the longitudinal and optical conductivities of graphene in the presence of a conducting plate until $2$-loop calculations and, since we computed them in terms of renormalized parameters (observable quantities), they could be interesting for experimental verifications.

%%%%%%%%%%%%%%%%%%%%%%%%%%%%%%%%%%%%%%%%%%%%%%%%%%%%%%%%%%%%%%%%%%%%%%%%%%
\section*{Acknowledgments}

The authors thank N. M. R. Peres, I. Sodemann, E. G. Mishchenko, E. C. Marino, A.
N. Braga, J. D. L. Silva, E. Granhen, W. Pires and L. Fernández for
fruitful discussions.
D.C.P. was supported by Coordenação de Aperfeiçoamento
de Pessoal de Nível Superior (CAPES/Brazil) -- Process 88882.445192/2018-01 and Process 88881.187657/2018-01.
D.T.A. was supported by UFPA via Licen\c{c}a Capacita\c{c}\~{a}o (Portaria 5603/2019), and thanks the hospitality of the University of Minho (Portugal), as well as that of the International Iberian Nanotechnology Laboratory (INL-Portugal).
V. S. A. is partially supported by Conselho Nacional de Desenvolvimento
Científico e Tecnlógico (CNPq) and by CAPES/NUFFIC, finance code 0112.

%%%%%%%%%%%%%%%%%%%%%%%%%%%%%%%%%%%%%%%%%%%%%%%%%%%%%%%%%%%%%%%%%%%%%%%%%%
\appendix

\section{The Sokhotski-Plemelj identity }\label{sec:The-Sokhotski-Plemelj-identity}

Usually, the Sokhotski-Plemelj identity is presented as \cite{galapon2016cauchy}
\begin{equation}
\lim_{\epsilon\rightarrow0}\int_{a}^{b}\frac{f(x)}{x-x_{0}\pm i\epsilon}dx=P\int_{a}^{b}\frac{f(x)}{x-x_{0}}dx\mp i\pi f(x_{0}),
\end{equation}
where $P$ is the Cauchy principal value. One can extend its definition to accomplish higher order poles \cite{galapon2016cauchy},
\begin{equation}
\lim_{\epsilon\rightarrow0}\int_{a}^{b}\frac{f(x)}{(x-x_{0}\pm i\epsilon)^{n+1}}dx=\#\int_{a}^{b}\frac{f(x)}{(x-x_{0})^{n+1}}dx\mp i\pi\frac{f^{(n)}(x_{0})}{n!},\label{eq:SP-identity}
\end{equation}
where $\#$ represents the Hadamard Finite-Part Integral (an extension of the Cauchy principal value integral), and is defined as
\begin{equation}
\#\int_{a}^{b}\frac{f(x)}{(x-x_{0})^{n+1}}dx=\lim_{\epsilon\rightarrow0}\left[\int_{a}^{x_{0}-\epsilon}\frac{f(x)}{(x-x_{0})^{n+1}}+\int_{x_{0}+\epsilon}^{b}\frac{f(x)}{(x-x_{0})^{n+1}}-H_{n}(x_{0},\epsilon)\right],\label{eq:real-part}
\end{equation}
where
\begin{equation}
H_{0}=0,
\end{equation}
and
\begin{equation}
H_{n}=\sum_{k=0}^{n-1}\frac{h^{(k)}(x_{0})}{k!(n-k)}\frac{(1-(-1)^{n-k})}{\epsilon^{n-k}},\qquad n=1,2,....\label{eq:H-func}
\end{equation}
Hereafter, we use the above representations to compute the real and imaginary parts of the $2$-loop corrections to the polarization tensor for $x_q=\omega/(v_F |\mathbf{q}|)>1$.

%%%%%%%%%%%%%%%%%%%%%%%%%%%%%%%%%%%%%%%%%%%%%%%%%%%%%%%%%%%%%%%%%%%%%%%%%
\subsection{$\Pi_{2a}$ diagram}\label{subsec:Pi2a-diagram}

First, we compute the real and imaginary parts of the integral $I_{a^{\prime\prime}}$ [Eq. (\ref{eq:Ib-pi2a})] of the $\Pi_{2a}$ diagram, for $x_q>1$.

%%%%%%%%%%%%%%%%%%%%%%%%%%%%%%%%%%%%%%%%%%%%%%%%%%%%%%%%%%%%%%%%%%%%%%%%%%%%%%%%%%%%%%%
\subsubsection{Real part}

Lets make a change of variables in Eq. (\ref{eq:Ib-pi2a}),
\begin{align}
w & =\cosh\mu,\qquad d\mu=\frac{dw}{\sqrt{w^{2}-1}},
\end{align}
obtaining
\begin{equation}
I_{a^{\prime\prime}}(\rho_{0}|\mathbf{q}|,y_q) =\int_{0}^{2\pi}d\nu\int_{1}^{\infty}dw\frac{H(w,y_q,\nu)}{(w-y_q)^{2}},
\end{equation}
where
\begin{equation}
H(w,y_q,\nu)=\frac{1}{\pi}\frac{\sin^{2}\nu(w-\cos\nu)(w^{2}+y_{q}^{2})}{(w+y_q)^{2}\sqrt{w^{2}-1}}F\left(\frac{\rho_{0}|\mathbf{q}|}{2}(w-\cos\nu),\Lambda\right).
\end{equation}
Then, from Eqs. (\ref{eq:real-part}) and (\ref{eq:H-func}), we find
the real part is given by
\begin{equation}
{\rm Re}\,[I_{a^{\prime\prime}}(\rho_{0}|\mathbf{q}|,x_q)]=\lim_{\epsilon\rightarrow0^{+}}\left[\int_{1}^{x_q-\epsilon}\frac{H(w,x_q,\nu)}{(w-x_q)^{2}}dw+\int_{x_q+\epsilon}^{\infty}\frac{H(w,x_q,\nu)}{(w-x_q)^{2}}dw-\frac{2H(x_q,x_q,\nu)}{\epsilon}\right].
\end{equation}

%%%%%%%%%%%%%%%%%%%%%%%%%%%%%%%%%%%%%%%%%%%%%%%%%%%%%%%%%%%%%%%%%%%%%%%%%%%%%
\subsubsection{Imaginary part}

From Eq. (\ref{eq:SP-identity}), we obtain the imaginary part:
\begin{equation}
{\rm Im}\,[I_{a^{\prime\prime}}(\rho_{0}|\mathbf{q}|,x_q)] =\int_{0}^{2\pi}d\nu\left.\frac{dH(w,x_q,\nu)}{dw}\right|_{w=x_q},
\end{equation}
where
\begin{align}
\left.\frac{dH(w,x_q,\nu)}{dw}\right|_{w=x_q} & =\frac{\sin^{2}\nu}{2\pi(x_{q}^{2}-1)^{3/2}}\left[(x_q\cos\nu-1)F\left(\frac{\rho_{0}|\mathbf{q}|}{2}\left(x_q-\cos\nu\right),\Lambda\right)+(x_{q}^{2}-1)(x_q-\cos\nu)\right.\nonumber \\
 & \left.\times \left.\frac{d}{dw}F\left(\frac{\rho_{0}\left|\mathbf{q}\right|}{2}(w-\cos\nu),\Lambda\right)\right|_{w=x_q}\right].
\end{align}

%%%%%%%%%%%%%%%%%%%%%%%%%%%%%%%%%%%%%%%%%%%%%%%%%%%%%%%%%%%%%%%%%%%%%%%%%%%%%%%%%%%%%%%
\subsection{$\Pi_{2b}$ diagram}\label{subsec:Pi2b-diagram}

Here, we compute the real and imaginary parts of the $\Pi_{2b}$ correction to the vacuum polarization, represented by Eq. (\ref{eq:p2b}). In the following steps, our calculations are similar to Ref. \cite{sodemann2012interaction}, though leading to different representations, they generate the same numerical results.

Since Eq. (\ref{eq:p2b}) has a first order pole, we can apply the
usual definition of the Sokhotski-Plemelj identity, but in terms of
the delta function \cite{sodemann2012interaction}:
\begin{equation}
\frac{1}{x-x_{0}\pm i\epsilon}=P\frac{1}{x-x_{0}}\mp i\pi\delta(x-x_{0}),
\end{equation}
which can be also generalized for a function on the denominator,
\begin{equation}
\frac{1}{g(x)\pm i\epsilon}=\sum_{i}\left[P\frac{1}{g(x_{i})}\mp i\pi\delta(g(x_{i}))\right],\label{eq:sokhotski-n1}
\end{equation}
where $g(x)$ is an invertible function in the region of integration, and the delta function of a function is given by \cite{snieder2004guided}
\begin{equation}
\delta(g(x))=\sum_{i}\frac{\delta(x-x_{i})}{|g^{\prime}(x_{i})|},
\end{equation}
assuming $x=x_{i}$ are the zeros of $g(x)$.

In Eq. (\ref{eq:p2b}), making $x_q=\cosh(\lambda/2)$ \cite{sodemann2012interaction}, we must define two functions in the denominator, namely
\begin{equation}
g_{+}(a,b,\lambda)=\cosh(a+b)-\cosh\lambda,
\end{equation}
and
\begin{equation}
g_{-}(a,b,\lambda)=\cosh(a-b)-\cosh\lambda,
\end{equation}
which have poles at $a_{\pm}=\lambda\mp b$. Hence, Eq. (\ref{eq:p2b}) leads to
\begin{equation}
p_{2b}(\rho_{0}|\mathbf{q}|,x_q)=-\frac{1}{16\pi^{3}}\int_{0}^{\infty}db\int_{b}^{\infty}da\frac{U(a,b,\rho_{0}|\mathbf{q}|,\lambda(x_q))}{[g_{+}(a,b,\lambda(x_q))-i\epsilon][g_{-}(a,b,\lambda(x_q))-i\epsilon]},\label{eq:p2b-modified}
\end{equation}
where
\begin{align}
U(a,b,\rho_0|\mathbf{q}|,\lambda)& =(\cosh^{2}a-\cosh^{2}b)I_{1}(a,b,\rho_0|\mathbf{q}|)+\left(\frac{\cosh\lambda+3}{2}\right)(\cosh a+\cosh b)I_{2}(a,b,\rho_0|\mathbf{q}|)  \nonumber \\
&
+\frac{\cosh\lambda+1}{2}(\cosh a-\cosh b)I_{3}(a,b,\rho_0|\mathbf{q}|).
\end{align}

Next, we explicit the real and imaginary parts of $p_{2b}$.

%%%%%%%%%%%%%%%%%%%%%%%%%%%%%%%%%%%%%%%%%%%%%%%%%%%%%%%%%%%%%%%%%%%%%%%%%%%%
\subsubsection{Real part}

As mentioned before, the real part of $p_{2b}$ will be calculated
by taking the principal value of (\ref{eq:p2b-modified}), namely
\begin{align}
&{\rm Re}[p_{2b}(\rho_{0}|\mathbf{q}|,x_q)] =-\frac{1}{16\pi^{3}}\lim_{\epsilon\rightarrow0^{+}}\int_{0}^{\infty}db\left[\int_{b}^{\lambda(x_q)-b-\epsilon}da\frac{U(a,b,\rho_{0}|\mathbf{q}|,\lambda(x_q))}{g_{+}(a,b,\lambda(x_q))g_{-}(a,b,\lambda(x_q))}\right.\nonumber \\
 & +\left.\int_{\lambda(x_q)-b+\epsilon}^{\lambda(x_q)+b-\epsilon}da\frac{U(a,b,\rho_{0}|\mathbf{q}|,\lambda(x_q))}{g_{+}(a,b,\lambda(x_q))g_{-}(a,b,\lambda(x_q))}+\int_{\lambda(x_q)+b+\epsilon}^{\infty}da\frac{U(a,b,\rho_{0}|\mathbf{q}|,\lambda(x_q))}{g_{+}(a,b,\lambda(x_q))g_{-}(a,b,\lambda(x_q))}\right].
\end{align}

%%%%%%%%%%%%%%%%%%%%%%%%%%%%%%%%%%%%%%%%%%%%%%%%%%%%%%%%%%%%%%%%%%%%%%
\subsubsection{Imaginary part}

The poles of the functions $g_{+}(a,b,\lambda)$ and $g_{-}(a,b,\lambda)$ are not the same. Therefore, we have
\begin{equation}
\delta(g_{+}(a,b,\lambda))=\frac{\delta(a-(\lambda-b))}{\sinh\lambda},
\end{equation}
\begin{equation}
\delta(g_{-}(a,b,\lambda))=\frac{\delta(a-(\lambda+b))}{\sinh\lambda}.
\end{equation}
Hence, the imaginary part of Eq. (\ref{eq:p2b-modified}) becomes
\begin{align}
{\rm Im}[p_{2b}(\rho_{0}|\mathbf{q}|,x_q)] & =-\frac{1}{32\pi^{2}\sinh\lambda(x_q)}\left\{ \int_{0}^{\lambda(x_q)/2}\frac{db}{\sinh b}\left[\frac{U(\lambda(x_q)+b,b,\lambda(x_q))}{\sinh(b+\lambda(x_q))}\right.\right.\nonumber \\
 & \left.\left.-\frac{U(\lambda(x_q)-b,b,\lambda(x_q))}{\sinh(\lambda(x_q)-b)}\right]+\int_{\lambda(x_q)/2}^{\infty}\frac{db}{\sinh b}\frac{U(\lambda(x_q)+b,b,\lambda(x_q))}{\sinh(b+\lambda(x_q))}\right\} .
\end{align}

%%%%%%%%%%%%%%%%%%%%%%%%%%%%%%%%%%%%%%%%%%%%%%%%%%%%%%%%%%%%%%%%%%%%%%
\section{The optical limit}\label{sec:opt-limit}

Obtaining $\sigma_{{\rm opt},2a}(\omega)$ requires the calculation $\Pi_{2a}$ in Eq. (\ref{eq-sigma-opt-2a}). Using Eq. (\ref{eq:Pi-2a-cartesian}), we have
\begin{equation}
\Pi_{2a}(\rho_{0},\omega,\mathbf{q}) =-\frac{Ne^{4}}{2\kappa}\int\frac{d^{2}k}{(2\pi)^{2}}\frac{\mathbf{k}\cdot(\mathbf{k}+\mathbf{q})-|\mathbf{k}||\mathbf{k}+\mathbf{q}|}{|\mathbf{k}+\mathbf{q}|}\frac{\left[v_{F}^{2}(|\mathbf{k}|+|\mathbf{k}+\mathbf{q}|)^{2}+\omega^{2}\right]}{\left[v_{F}^{2}(|\mathbf{k}|+|\mathbf{k}+\mathbf{q}|)^{2}-\omega^{2}\right]^{2}}
 \left[\ln\left(\Lambda/|\mathbf{k}|\right)-F(\rho_{0}|\mathbf{k}|)\right],
\end{equation}
where we made $\mathbf{k}\rightarrow-\mathbf{k}-\mathbf{q}$. Expanding until order $|\mathbf{q}|^{2}$ leads to
\begin{align}
\Pi_{2a}(\rho_{0},\omega,\mathbf{q}) & =-\frac{Ne^{4}}{2\kappa}\int\frac{d^{2}k}{(2\pi)^{2}}\frac{\frac{1}{2}|\mathbf{q}|^{2}\left(\cos^{2}\theta-1\right)}{|\mathbf{k}|}\left(\frac{4v_{F}^{2}|\mathbf{k}|^{2}-q_{0}^{2}}{4v_{F}^{2}|\mathbf{k}|^{2}+q_{0}^{2}}\right)\left[\ln\left(\Lambda/|\mathbf{k}|\right)-F(\rho_{0}|\mathbf{k}|)\right],
\end{align}
where $\theta$ is the angle between $\mathbf{k}$ and $\mathbf{q}$. Then, integrating in the polar coordinate system, we get
\begin{align}
\Pi_{2a}(\rho_{0},\omega,\mathbf{q}) & =\frac{e^{2}|\mathbf{q}|^{2}v_{F}\alpha}{8\pi}\int d|\mathbf{k}|\frac{\left(4v_{F}^{2}|\mathbf{k}|^{2}+\omega^{2}\right)}{\left(4v_{F}^{2}|\mathbf{k}|^{2}-\omega^{2}\right)^{2}}\left[\ln\left(\frac{\Lambda}{|\mathbf{k}|}\right)-F(\rho_{0}|\mathbf{k}|)\right].
\end{align}
The above integral has a second order pole, and its imaginary part will be obtained from the following formula, better explained in \ref{sec:The-Sokhotski-Plemelj-identity}:
\begin{equation}
{\rm Im}\left\{ \lim_{\epsilon\rightarrow0}\int_{a}^{b}\frac{f(x)}{(x-x_{0}\pm i\epsilon)^{n+1}}dx\right\} =\mp\pi\frac{f^{(n)}(x_{0})}{n!}.
\end{equation}
In our case, $n=1$ and we define
\begin{equation}
f(|\mathbf{k}|)=\frac{\left(4v_{F}^{2}|\mathbf{k}|^{2}+\omega^{2}\right)\left[\ln\left(\frac{\Lambda}{|\mathbf{k}|}\right)-F(\rho_{0}|\mathbf{k}|)\right]}{\left(2v_{F}|\mathbf{k}|+\omega\right)^{2}},
\end{equation}
with derivative given by
\begin{equation}
\frac{df}{d|\mathbf{k}|}=-\frac{v_{F}}{\omega}-\frac{\rho_{0}}{2}\left.\frac{dF(\rho_{0}|\mathbf{k}|)}{d(\rho_{0}|\mathbf{k}|)}\right|_{|\mathbf{k}|=\omega/2v_{F}}.
\end{equation}
Therefore, we have that
\begin{equation}
{\rm Im}\left[2\Pi_{2a}(\rho_{0},\omega,\mathbf{q})\right]=-\frac{e^{2}}{4}\frac{|\mathbf{q}|^{2}}{\omega}\frac{\alpha}{4}\left(1+\frac{\rho_{0}\omega}{v_{F}}\frac{1}{2}\left.\frac{dF(\rho_{0}|\mathbf{k}|)}{d(\rho_{0}|\mathbf{k}|)}\right|_{|\mathbf{k}|=\omega/2v_{F}}\right).
\end{equation}
Hence, from (\ref{formula-kubo}), the above contribution to the real part of the conductivity will be
\begin{equation}
{\rm Re}[\sigma_{{\rm opt},2a}(\rho_{0},\omega)]=\sigma_{0}\frac{\alpha}{4}\left(1+\frac{\rho_{0}\omega}{v_{F}}\frac{1}{2}\left.\frac{dF(\rho_{0}|\mathbf{k}|)}{d(\rho_{0}|\mathbf{k}|)}\right|_{|\mathbf{k}|=\omega/2v_{F}}\right),
\label{eq-sigma-opt-2a-sol} 
\end{equation}
where the first term was obtained in \cite{mishchenko2008minimal}.

The $\sigma_{{\rm opt},2b}(\omega)$ contribution is given by Eq. (\ref{eq-sigma-opt-2b}). Expanding the $\Pi_{2b}$ contribution until order $|\mathbf{q}|^{2}$ in Eq. (\ref{PI-V-2}), it follows that:
\begin{align}
\Pi_{2b}(\rho_{0},\omega,\mathbf{q}) & =-\frac{Ne^{4}|\mathbf{q}|^{2}}{2\kappa(2\pi)^{3}}\int_{0}^{\infty}d|\mathbf{k}|\int_{0}^{\infty}d|\mathbf{p}|\int_{0}^{2\pi}d\theta_{k}\int_{0}^{2\pi}d\theta_{p}|\mathbf{k}||\mathbf{p}|\left[\cos(\theta_{k}-\theta_{p})-\cos(\theta_{k}+\theta_{p})\right]\nonumber \\
& \times\frac{\left[\frac{\omega^{2}}{|\mathbf{k}||\mathbf{p}|}\cos(\theta_{k}-\theta_{p})+4v_{F}^{2}\right]\left(1-e^{-2\rho_{0}|\mathbf{k}-\mathbf{p}|}\right)}{|\mathbf{k}-\mathbf{p}|\left(4v_{F}^{2}|\mathbf{k}|^{2}-\omega^{2}\right)\left(4v_{F}^{2}|\mathbf{p}|^{2}-\omega^{2}\right)},
\end{align}
where we have made $\mathbf{q}=(|\mathbf{q}|,0)$ such that $\theta_{kq}=\theta_{k}$ and $\theta_{pq}=\theta_{p}$. From a change of variables, $\theta=\theta_{k}-\theta_{p}$ and $\varphi=\theta_{k}+\theta_{p}$, one can easily obtain that
\begin{align}
\Pi_{2b}(\rho_{0},\omega,\mathbf{q}) & =-\frac{e^{4}}{2\pi\kappa}|\mathbf{q}|^{2}\int_{0}^{\pi}\frac{d\theta}{\pi}\int_{0}^{\infty}d|\mathbf{k}|\int_{0}^{\infty}d|\mathbf{p}||\mathbf{k}||\mathbf{p}|\cos\theta\nonumber \\
& \times\frac{\left(\frac{\omega^{2}}{|\mathbf{k}||\mathbf{p}|}\cos\theta+4v_{F}^{2}\right)\left(1-e^{-2\rho_{0}\sqrt{|\mathbf{k}|^{2}+|\mathbf{p}|^{2}-2|\mathbf{k}||\mathbf{p}|\cos\theta}}\right)}{\left(4v_{F}^{2}|\mathbf{k}|^{2}-\omega^{2}\right)\left(4v_{F}^{2}|\mathbf{p}|^{2}-\omega^{2}\right)\sqrt{|\mathbf{k}|^{2}+|\mathbf{p}|^{2}-2|\mathbf{k}||\mathbf{p}|\cos\theta}}.\label{eq:pi2b-q2}
\end{align}
From Eq. (\ref{eq:sokhotski-n1}), we find that the imaginary part of the above equation will be given in terms of
\begin{equation}
{\rm Im}\left[\frac{1}{4v_{F}^{2}|\mathbf{p}|^{2}-\omega^{2}}\right]=\frac{\pi}{4\omega v_{F}}\delta(|\mathbf{p}|-\omega/2v_{F})\qquad\text{and}\qquad{\rm Im}\left[\frac{1}{4v_{F}^{2}|\mathbf{k}|^{2}-\omega^{2}}\right]=\frac{\pi}{4\omega v_{F}}\delta(|\mathbf{k}|-\omega/2v_{F}).
\end{equation}
Due to symmetry, we can multiply the integral in (\ref{eq:pi2b-q2}) by $2$ and consider only the imaginary part of \textbf{$\mathbf{p}$}, giving ($u=2v_{F}|\mathbf{k}|/\omega$)
\begin{equation}
{\rm Im}[\Pi_{2b}(\rho_{0},\omega,\mathbf{q})]=\frac{\sigma_{0}|\mathbf{q}|^{2}\alpha}{\omega}\int_{0}^{\pi}\frac{d\theta}{\pi}\int_{0}^{\infty}du\frac{\cos\theta(u+\cos\theta)\left[1-\exp\left(-\frac{\rho_{0}\omega}{v_{F}}\sqrt{u^{2}+1-2u\cos\theta}\right)\right]}{\left(1-u^{2}\right)\sqrt{u^{2}+1-2u\cos\theta}},
\end{equation}
where the first term in the integral was determined in \cite{mishchenko2008minimal}, giving
\begin{equation}
{\rm Re} 
[\sigma_{{\rm opt},2b}(\rho_0,\omega)]=\sigma_{0}\alpha\left[\frac{8-3\pi}{6}+\int_{0}^{\pi}\frac{d\theta}{\pi}\int_{0}^{\infty}du\frac{\cos\theta(u+\cos\theta)\exp\left(-\frac{\rho_{0}\omega}{v_{F}}\sqrt{u^{2}+1-2u\cos\theta}\right)}{\left(1-u^{2}\right)\sqrt{u^{2}+1-2u\cos\theta}}\right].
\label{eq-sigma-opt-2b-sol} 
\end{equation}

\bibliography{refs}

\begin{thebibliography}{10}
\expandafter\ifx\csname url\endcsname\relax
  \def\url#1{\texttt{#1}}\fi
\expandafter\ifx\csname urlprefix\endcsname\relax\def\urlprefix{URL }\fi
\expandafter\ifx\csname href\endcsname\relax
  \def\href#1#2{#2} \def\path#1{#1}\fi

\bibitem{roberts1994dyson}
C.~D. Roberts, A.~G. Williams, Dyson-{S}chwinger equations and their
  application to hadronic physics, Progress in Particle and Nuclear Physics 33
  (1994) 477.
\newblock \href {https://doi.org/10.1016/0146-6410(94)90049-3}
  {\path{doi:10.1016/0146-6410(94)90049-3}}.

\bibitem{marino1993quantum}
E.~Marino, Quantum electrodynamics of particles on a plane and the
  {Chern-Simons} theory, Nuclear Physics B 408~(3) (1993) 551.
\newblock \href {https://doi.org/10.1016/0550-3213(93)90379-4}
  {\path{doi:10.1016/0550-3213(93)90379-4}}.

\bibitem{alves2013chiral}
V.~S. Alves, W.~S. Elias, L.~O. Nascimento, V.~Juri{\v{c}}i{\'c}, F.~Pe{\~n}a,
  Chiral symmetry breaking in the pseudo-quantum electrodynamics, Physical
  Review D 87~(12) (2013) 125002.
\newblock \href {https://doi.org/10.1103/PhysRevD.87.125002}
  {\path{doi:10.1103/PhysRevD.87.125002}}.

\bibitem{marino2015interaction}
E.~Marino, L.~O. Nascimento, V.~S. Alves, C.~M. Smith, Interaction induced
  quantum valley {H}all effect in graphene, Physical Review X 5~(1) (2015)
  011040.
\newblock \href {https://doi.org/10.1103/PhysRevX.5.011040}
  {\path{doi:10.1103/PhysRevX.5.011040}}.

\bibitem{nascimento2015chiral}
L.~O. Nascimento, V.~S. Alves, F.~Pe{\~n}a, C.~M. Smith, E.~Marino,
  Chiral-symmetry breaking in pseudoquantum electrodynamics at finite
  temperature, Physical Review D 92~(2) (2015) 025018.
\newblock \href {https://doi.org/10.1103/PhysRevD.92.025018}
  {\path{doi:10.1103/PhysRevD.92.025018}}.

\bibitem{menezes2016fermi}
N.~Menezes, V.~S. Alves, C.~de~Morais~Smith, Fermi-velocity renormalization due
  to interactions in graphene: the influence of a weak magnetic field, Eur.
  Phys. JB 89~(271) (2016).
\newblock \href {https://doi.org/10.1140/epjb/e2016-70606-4}
  {\path{doi:10.1140/epjb/e2016-70606-4}}.

\bibitem{menezes2016influence}
N.~Menezes, V.~S. Alves, C.~M. Smith, The influence of a weak magnetic field in
  the renormalization-group functions of (2+ 1)-dimensional {D}irac systems,
  The European Physical Journal B 89~(12) (2016) 271.
\newblock \href {https://doi.org/10.1140/epjb/e2016-70606-4}
  {\path{doi:10.1140/epjb/e2016-70606-4}}.

\bibitem{alves2017dynamical}
V.~S. Alves, R.~O. Junior, E.~Marino, L.~O. Nascimento, Dynamical mass
  generation in pseudoquantum electrodynamics with four-fermion interactions,
  Physical Review D 96~(3) (2017) 034005.
\newblock \href {https://doi.org/10.1103/PhysRevD.96.034005}
  {\path{doi:10.1103/PhysRevD.96.034005}}.

\bibitem{menezes2017spin}
N.~Menezes, V.~S. Alves, E.~Marino, L.~Nascimento, L.~O. Nascimento, C.~M.
  Smith, Spin g-factor due to electronic interactions in graphene, Physical
  Review B 95~(24) (2017) 245138.
\newblock \href {https://doi.org/10.1103/PhysRevB.95.245138}
  {\path{doi:10.1103/PhysRevB.95.245138}}.

\bibitem{nascimento2017introduction}
L.~O. Nascimento, Introduction to topological phases and electronic
  interactions in (2+ 1) dimensions, Brazilian Journal of Physics 47~(2) (2017)
  215--230.
\newblock \href {https://doi.org/10.1007/s13538-017-0485-0}
  {\path{doi:10.1007/s13538-017-0485-0}}.

\bibitem{silva2017inhibition}
J.~D.~L. Silva, A.~N. Braga, W.~P. Pires, V.~S. Alves, D.~T. Alves, E.~Marino,
  Inhibition of the {F}ermi velocity renormalization in a graphene sheet by the
  presence of a conducting plate, Nuclear Physics B 920 (2017) 221.
\newblock \href {https://doi.org/10.1016/j.nuclphysb.2017.04.014}
  {\path{doi:10.1016/j.nuclphysb.2017.04.014}}.

\bibitem{marino2018screening}
E.~Marino, D.~Niemeyer, V.~S. Alves, T.~H. Hansson, S.~Moroz, Screening and
  topological order in thin superconducting films, New Journal of Physics
  20~(8) (2018) 083049.
\newblock \href {https://doi.org/10.1088/1367-2630/aadb36}
  {\path{doi:10.1088/1367-2630/aadb36}}.

\bibitem{marino2018quantum}
E.~Marino, L.~O. Nascimento, V.~S. Alves, N.~Menezes, C.~M. Smith,
  Quantum-electrodynamical approach to the exciton spectrum in transition-metal
  dichalcogenides, 2D Materials 5~(4) (2018) 041006.
\newblock \href {https://doi.org/10.1088/2053-1583/aacc3f}
  {\path{doi:10.1088/2053-1583/aacc3f}}.

\bibitem{alves2018two}
V.~S. Alves, T.~Macri, G.~C. Magalh{\~a}es, E.~Marino, L.~O. Nascimento,
  Two-dimensional {Y}ukawa interactions from nonlocal {P}roca quantum
  electrodynamics, Physical Review D 97~(9) (2018) 096003.
\newblock \href {https://doi.org/10.1103/PhysRevD.97.096003}
  {\path{doi:10.1103/PhysRevD.97.096003}}.

\bibitem{pires2018cavity}
W.~P. Pires, J.~D.~L. Silva, A.~N. Braga, V.~S. Alves, D.~T. Alves, E.~Marino,
  Cavity effects on the {F}ermi velocity renormalization in a graphene sheet,
  Nuclear Physics B 932 (2018) 529.
\newblock \href {https://doi.org/10.1016/j.nuclphysb.2018.05.010}
  {\path{doi:10.1016/j.nuclphysb.2018.05.010}}.

\bibitem{marino2020graphene}
E.~C. Marino, From graphene to quantum computation: An expedition to the
  {D}irac sea, in: Strongly Coupled Field Theories for Condensed Matter and
  Quantum Information Theory, Springer, 2020, pp. 339--353.
\newblock \href {https://doi.org/10.1007/978-3-030-35473-2_15}
  {\path{doi:10.1007/978-3-030-35473-2_15}}.

\bibitem{magalhaes2020pseudo}
G.~C. Magalh{\~a}es, V.~S. Alves, E.~C. Marino, L.~O. Nascimento, Pseudo
  quantum electrodynamics and {Chern-Simons} theory coupled to two-dimensional
  electrons, Physical Review D 101~(11) (2020) 116005.
\newblock \href {https://doi.org/10.1103/PhysRevD.101.116005}
  {\path{doi:10.1103/PhysRevD.101.116005}}.

\bibitem{zhang1998electromagnetic}
K.~Zhang, D.~Li, K.~Chang, K.~Zhang, D.~Li, Electromagnetic theory for
  microwaves and optoelectronics, Springer, 1998.

\bibitem{barnes2014effective}
E.~Barnes, E.~Hwang, R.~Throckmorton, S.~D. Sarma, Effective field theory,
  three-loop perturbative expansion, and their experimental implications in
  graphene many-body effects, Physical Review B 89~(23) (2014) 235431.
\newblock \href {https://doi.org/10.1103/PhysRevB.89.235431}
  {\path{doi:10.1103/PhysRevB.89.235431}}.

\bibitem{gonzalez1994non}
J.~Gonz{\'a}lez, F.~Guinea, M.~Vozmediano, Non-fermi liquid behavior of
  electrons in the half-filled honeycomb lattice (a renormalization group
  approach), Nuclear Physics B 424~(3) (1994) 595.
\newblock \href {https://doi.org/10.1016/0550-3213(94)90410-3}
  {\path{doi:10.1016/0550-3213(94)90410-3}}.

\bibitem{neto2009electronic}
A.~C. Neto, F.~Guinea, N.~M. Peres, K.~S. Novoselov, A.~K. Geim, The electronic
  properties of graphene, Reviews of modern physics 81~(1) (2009) 109.
\newblock \href {https://doi.org/10.1103/RevModPhys.81.109}
  {\path{doi:10.1103/RevModPhys.81.109}}.

\bibitem{de2012space}
F.~de~Juan, M.~Sturla, M.~A. Vozmediano, Space dependent {F}ermi velocity in
  strained graphene, Physical review letters 108~(22) (2012) 227205.
\newblock \href {https://doi.org/10.1103/PhysRevLett.108.227205}
  {\path{doi:10.1103/PhysRevLett.108.227205}}.

\bibitem{raoux2010velocity}
A.~Raoux, M.~Polini, R.~Asgari, A.~Hamilton, R.~Fazio, A.~H. MacDonald,
  Velocity-modulation control of electron-wave propagation in graphene,
  Physical Review B 81~(7) (2010) 073407.
\newblock \href {https://doi.org/10.1103/PhysRevB.81.073407}
  {\path{doi:10.1103/PhysRevB.81.073407}}.

\bibitem{stauber2017interacting}
T.~Stauber, P.~Parida, M.~Trushin, M.~V. Ulybyshev, D.~L. Boyda, J.~Schliemann,
  Interacting electrons in graphene: {F}ermi velocity renormalization and
  optical response, Physical review letters 118~(26) (2017) 266801.
\newblock \href {https://doi.org/10.1103/PhysRevLett.118.266801}
  {\path{doi:10.1103/PhysRevLett.118.266801}}.

\bibitem{mahan2013many}
G.~D. Mahan, Many-particle physics, Springer Science \& Business Media, 2013.

\bibitem{sodemann2012interaction}
I.~Sodemann, M.~M. Fogler, Interaction corrections to the polarization function
  of graphene, Physical Review B 86~(11) (2012) 115408.
\newblock \href {https://doi.org/10.1103/PhysRevB.86.115408}
  {\path{doi:10.1103/PhysRevB.86.115408}}.

\bibitem{kotov2008electron}
V.~N. Kotov, B.~Uchoa, A.~C. Neto, Electron-electron interactions in the vacuum
  polarization of graphene, Physical Review B 78~(3) (2008) 035119.
\newblock \href {https://doi.org/10.1103/PhysRevB.78.035119}
  {\path{doi:10.1103/PhysRevB.78.035119}}.

\bibitem{pisarski1984chiral}
R.~D. Pisarski, Chiral-symmetry breaking in three-dimensional electrodynamics,
  Physical Review D 29~(10) (1984) 2423.
\newblock \href {https://doi.org/10.1063/1.4943300}
  {\path{doi:10.1063/1.4943300}}.

\bibitem{appelquist1985chiral}
T.~W. Appelquist, Chiral symmetry breaking in quantum field theory, Progress of
  Theoretical Physics Supplement 85 (1985) 244--249.
\newblock \href {https://doi.org/10.1143/PTP.85.244}
  {\path{doi:10.1143/PTP.85.244}}.

\bibitem{galapon2016cauchy}
E.~A. Galapon, The {C}auchy principal value and the {H}adamard finite part
  integral as values of absolutely convergent integrals, Journal of
  Mathematical Physics 57~(3) (2016) 033502.

\bibitem{mishchenko2008minimal}
E.~Mishchenko, Minimal conductivity in graphene: Interaction corrections and
  ultraviolet anomaly, EPL (Europhysics Letters) 83~(1) (2008) 17005.
\newblock \href {https://doi.org/10.1209/0295-5075/83/17005}
  {\path{doi:10.1209/0295-5075/83/17005}}.

\bibitem{snieder2004guided}
R.~Snieder, C.~U. Press, A Guided Tour of Mathematical Methods: For the
  Physical Sciences, A Guided Tour of Mathematical Methods for the Physical
  Sciences, Cambridge University Press, 2004.
\newblock \href {https://doi.org/10.1007/s13538-017-0485-0}
  {\path{doi:10.1007/s13538-017-0485-0}}.

\end{thebibliography}

\end{document}